\documentclass[a4paper,fleqn,usenatbib,useAMS]{mnras}

\usepackage{graphicx}	
\usepackage{amsmath}	
\usepackage{amssymb}	
\usepackage{multicol}        
\usepackage{bm}		
\usepackage{pdflscape}	
\usepackage{rotating}

\def\apjl{\rmfamily{ApJ~}}        
\def\apjs{\rmfamily{ApJS~}}       

\def\aap{\rmfamily{A\&A~}}        

\def\ao{\rmfamily{Appl.~Opt.~}}   

\newcommand\nh{ N_{\rm H}}

\newcommand\ks{\, \rm ks}

\newcommand\rg{r_{\rm g}}

\def\suzaku{{\it Suzaku~}}
\def\suzs{{\it Suzaku}}

\def\swift{{\it Swift~}}

\def\xmm{{\it XMM-Newton~}}
\def\xmms{{\it XMM-Newton}}

\def\nustar{{\it NuSTAR~}}
\def\nus{{\it NuSTAR}}
\def\kepler{{\it Kepler~}}
\def\bat{{\it BAT~}}

\def\A{{\rm\thinspace \AA}~}
\def\zw229{{Zw~229.015}}
\def\mk335{{Mrk~335}}
\def\mrk530{{Mrk~530}}
\def\IZw1{{I~Zw1}} 
\def\Akn120{{Akn~120}}
\def\ngc4051{{NGC~4051}}
\def\Ngc5548{{NGC~5548}}

\def\arcs{{\hbox{$^{\prime\prime}$~}}}
\def\deg{^{\circ}}
\def\A{{\rm\thinspace \AA}~}
\def\cm{{\rm\thinspace cm}}
\def\erg{{\rm\thinspace erg}}

\def\ga{{\rm\thinspace gauss}}

\def\keV{{\rm\thinspace keV}}

\def\Mpc{{\rm\thinspace Mpc}}

\def\s{{\rm\thinspace s}}
\def\ks{{\rm\thinspace ks}}

\def\ergcmps{\hbox{$\erg\cm\s^{-1}\,$}}

\usepackage{graphicx}

\title[X-ray characteristics of \zw229]{The nature of the soft-excess and spectral variability in the Seyfert 1 galaxy \zw229}

\author[Tripathi et al.]{S. Tripathi$^1$\thanks{tripathi@ap.smu.ca}, S. G. H. Waddell$^1$, L. C. Gallo$^1$, W. F. Welsh$^2$ and C-Y. Chiang$^3$\\ 
$^1$ Department of Astronomy and Physics, Saint Mary's University, Halifax, Canada \\
$^2$ Department of Astronomy, San Diego State University, 5500 Campanile Drive, San Diego, CA 92182, USA \\
$^3$ Department of Physics and Astronomy, Wayne State University, Detroit, MI 48201, USA\\
}

\date{\today}
\begin{document}




\maketitle
\label{firstpage}
\begin{abstract}
We have carried out a systematic analysis of the nearby (z=0.0279) active galaxy \zw229 using multi-epoch, multi-instrument and deep pointed observations with \xmms, \suzs, \swift and \nus.  Spectral and temporal variability are examined in detail on both the long (weeks-to-years) and short (hours) timescales. A deep \suzaku observation of the source shows two distinct spectral states; a bright-soft state and a dim-hard state in which changes in the power law component account for the differences. Partial covering, blurred reflection and soft Comptonisation models describe the X-ray spectra comparably well, but the smooth, rather featureless, spectrum may be favouring the soft Comptonisation scenario. Moreover, independent of the spectral model, the observed spectral variability is ascribed to the changes in the power law continuum only and do not require changes in the properties of the absorber or blurred reflector incorporated in the other scenarios. The multi-epoch observations between 2009 and 2018 can be described in similar fashion. This could be understood if the primary emission is originating at a large distance from a standard accretion disc or if the disc is optically thin and geometrically thick  as recently proposed for \zw229. Our investigation shows that \zw229 behaves similar to sources like \Akn120 and \mrk530, that exhibit a strong soft-excess, but weak Compton hump and Fe~K$\alpha$ emission.

 \end{abstract}

\begin{keywords}
X-ray: galaxies --
galaxies: active -- 
galaxies: nuclei -- 
galaxies: Seyfert -- \\
galaxies: individual: \zw229\ 
\end{keywords}
\section{INTRODUCTION}
Active Galactic Nuclei (AGN) display luminous and energetic central regions due to the presence of an actively accreting supermassive black hole at the core of the galaxy. The resulting emission covers the entire range of the electromagnetic spectrum ranging from the radio wavelengths, infra-red, optical, UV, X-rays up to Gamma-rays. Observational differences due to the orientation of observer towards the central engine and the observed luminosity in different wavelengths segregates AGN into different classes. Seyfert 1 galaxies are a subclass of AGN with an unobstructed line of sight to the innermost regions of the central engine \citep{Antonucci1993, Urry+1995} such that the line of sight does not intercept the molecular torus. 

Many AGN share the common property of copious and variable X-ray emission that emanates from the innermost regions surrounding the supermassive black hole. The physical mechanism responsible for the direct primary X-ray continuum is widely understood to be the inverse Compton scattering of seed ultraviolet photons (thermally produced by the accretion disk), by a corona of optically thin, thermal electrons above the disk \citep{Haardt+1991, Haardt+1993}. Moreover, observations also show a prominent `soft excess' component below $\sim$ 1 keV over the extrapolated coronal power law in the spectra for most Seyfert 1 AGN \citep{Pravdo+1981, Singh+1985, Arnaud+1985}, which is not well understood \citep{Garcia+2019, Middei+2019, Petrucci+2018}. To explain and reproduce the typical X-ray spectra in AGN, the primary power law X-ray continuum must also undergo modification via different processes including absorption by neutral or ionized gas, blurred reflection from the accretion disc or more distant reflection attributed to the cold molecular torus.  

Apart from the observed complexity in the spectra, AGN can also exhibit significant spectral and temporal variability over the short (hours-days) (e.g. \citealt{Ponti+2012, Chiang+2015}) and long (weeks-years) (e.g. \citealt{devries+2003,Vagnetti+2011}) time scales in terms of the flux and the spectral shape. Usually there is a correlation between the spectrum and the source intensity. 

While some AGN exhibit mild variability, a few of them exhibit extreme low states. Observations indicate that low flux states, in general, show relatively strong soft excess ($\leq$ 2 keV) and strong spectral complexity in the energy range 2-10 keV (e.g. \citealt{Gallo2006, Grupe+2008}). Previous studies have shown that the transition of AGN in low states is interesting as it provides the best clues to study the nature of the accretion disc, the accompanying relativistic features in detail, and the origin of variability \citep{Gallo+2007, Gallo+2011, Gallo+2015, Miller2007, Turner+2009}, although challenging to work with due to suppressed signal to noise ratio. This is even more interesting if the source has been observed simultaneously with multiple instruments and satellites (for instance, the simultaneous UV and X-ray observations can reliably quantify the X-ray weakness with $\alpha_{ox}$ measurements). Systematic spectral studies through plausible physical models in combination with the timing/variability studies serve as inevitably useful tools to retrieve reliable and meaningful information from the ``single-to-multiple" epoch/satellite observations of AGN.

\zw229 (CGCG 229-015, KIC 006932990, Zw~229-15) is a nearby ($z=0.0279$) Seyfert 1 AGN with the reverberation mapped black hole mass of $M_{BH}\sim10^{7}$ $M_{\odot}$ \citep{Barth+2011} and a moderate value of Galactic absorption column density $N_{H}~\sim~5.83~\times~10^{20}$ $cm^{-2}$ \citep{Kalberla+2005}. A considerable amount of work has been done on this source in the past decade as a result of being in the \kepler field of view. \zw229 is one of the AGN with the most complete, continuous, high quality and high resolution \kepler light curves that enabled the numerous studies on its optical variability properties \citep{Mushotzky+2011, Carini+2012, Edelson+2014, Kasliwal+2015, Kasliwal+2017, Adegoke+2017, Dobrotka+2017}.

The optical lightcurves of \zw229 obtained by \kepler show its highly variable character on multiple timescales. The source shows discrete variations of small amplitude ($\sim$~0.5$\%$) on timescales as short as tens of hours. The power spectral density (PSD) studies reported the discovery of the characteristic break timescale of ~5 days \citep{Edelson+2014}, whereas previous PSD studies presented the evidence of either no break \citep{Mushotzky+2011, Kelly+2014} or a possible break timescale of $\sim$ 90 days or $\sim$ 40 days \citep{Carini+2012}. Recently, \citet{Smith+2018} confirmed the already known 5-day break of \zw229 with greater significance than reported by previous studies.

\citet{Dobrotka+2017} examined the flickering activity of \zw229 in detail using \kepler and \swift observations. With \kepler data, they deduced different multiple component PSDs with low and high break frequencies as compared to the findings by \citet{Edelson+2014} and also found indications of presence of these break frequencies in the \swift XRT data. They found an explanation for these breaks consistent with the scenario in which the disc is optically thin and geometrically thick in the central region.

\citet{Adegoke+2017} studied the \xmm observation and found that both thermal Comptonisation and blurred reflection were statistically comparable to explain the origin of soft excess. With variability studies, they found lags in the soft and the hard X-ray energy band that correspond to two possible regions of X-ray emission within the system favouring thermal Comptonisation scenario. However, contemporaneous \suzaku and the recent \nustar observations have never been studied in detail before to systematically characterize the X-ray properties and spectral variability behaviour.

In this work, we employ the multiple archival observations with \xmms, \suzaku and \nustar to examine the spectral and temporal variability of \zw229 over various epochs. The details of the observations and data reduction are described in the next section. In Section~3, the deep \suzaku observation is examined.  The multi-epoch analysis is presented in Section 4.
Discussion and conclusions follow in Section 5 and 6, respectively.

\section{Observations \& Data Reduction}
\zw229 has been the focus of targeted \xmm \citep{Jansen+2001}, \suzaku \citep{Mitsuda+2007} and recent \nustar \citep{Harrison+2013} observations. Additionally, the AGN has also been observed in ultraviolet (UVOT; \citet{Roming+2005}) and X-ray (XRT; \citealt{Burrows+2005}, \bat; \citealt{Barthelmy+2005}) with \swift \citep{Gehrels+2004}. In this section we describe the reduction procedure for the archival data used in our work. A brief summary of the observations are presented in Table \ref{tab:datalog}. 

\subsection{\xmm}
The EPIC PN \citep{Struder+2001} camera was operated in the Prime Large window mode with the medium filter for the $\sim$ 30~ks duration of the \zw229 observation. We analyzed the Observation Data File (ODF) downloaded from the \xmm archive using the \xmm Science Analysis System (\texttt{SAS v15.0.0}) and the updated calibration files as of March 26, 2018. \newline
\indent We processed the EPIC PN data using task \texttt{EPPROC} and filtered them using the standard filtering criterion. For the spectral extraction, high background flaring was removed by creating a good time interval (gti) file above 10 keV for the full field with $<$0.5 cts s$^{-1}$ using the task \texttt{TABGTIGEN}. We checked for the photon pile-up by examining the data using the task \texttt{EPATPLOT} and found it absent in the observation. \newline
\indent Source photons were extracted from a 35\arcs circular region centred on the source and the background photons were extracted from a 35\arcs circular off-source region. EPIC response matrices were created using the \texttt{SAS} tasks \texttt{ARFGEN} and \texttt{RMFGEN}. \newline \indent We extracted background-subtracted, deadtime and vignetting corrected source lightcurve in the desired energy ranges using the task \texttt{EPICLCCORR}. \newline \indent MOS data were also reduced and found consistent with those taken by the EPIC PN camera. However, to ease comparison between different missions, we present only the PN spectrum in the analysis. We use the energy range 0.3-10.0 keV for all the spectral fits.\newline
\indent The Optical Monitor (OM) \citep{Mason+2001} was simultaneously operated in the imaging mode with the UVW1 and UVW2 filters. We processed the OM data using the task \texttt{OMICHAIN}. The count rates for the corresponding filters were obtained by specifying the RA and DEC of the source in the combined source list file. These were then converted into flux with the help of standard tables. We also corrected these fluxes for the Galactic extinction by following \citet{Fitzpatrick1999} reddening law with $R_{v}=3.1$  \newline
\indent We processed the Reflection Grating Spectrometer (RGS; \citet{denHerder+2001}) data with the task \texttt{RGSPROC}. The response files were produced using the task \texttt{RGSRMFGEN}. We checked the combined RGS spectra for the signatures of possible spectral features but the data quality is limited with the relatively short exposure. 

\subsection{\suzaku}
\zw229 was observed by \suzaku in the XIS-nominal position, with the two front-illuminated (XIS0 and XIS3) CCDs, the back-illuminated CCD (XIS1) and the HXD-PIN detector. We used HEASOFT, version 6.24 software and the recent calibration files as of June 7, 2016 and September 13, 2011 for XIS and PIN respectively to process the data.\newline
Cleaned event files were processed by the task \texttt{AEPIPELINE} and later used to extract data products using \texttt{XSELECT}. For each XIS instrument, source photons were extracted from a $~$230\arcs circular region centred on the source and background photons were extracted from three $\sim$120\arcs circular off-source regions avoiding the calibration sources. Response files were generated using the tasks \texttt{XISRMFGEN} and \texttt{XISSIMARFGEN}. \newline
\indent After checking that the XIS0 and XIS3 data were consistent with each other, the two were then combined to create a single spectrum using the task \texttt{ADDASCASPEC}. We also found the XIS1 data to be consistent with the front-illuminated (XIS0 and XIS3) detectors, though for the sake of simplicity, we used only the XIS0+XIS3 combined data. \\
\indent The XIS energy range 0.7-10.0 keV is used to determine the spectral fits. To avoid regions of poorly understood response, we ignore the spectral energy ranges 1.72-1.88 keV, 2.19-2.37 keV respectively \citep{Nowak+2011} before performing the spectral fits. \newline 
\indent The HXD-PIN spectrum was extracted using the tool \texttt{HXDPINXBPI} from the PIN cleaned events and the pseudo event lists were generated by task \texttt{AEPIPELINE}. For the non-imaging HXD PIN data, the background estimation needs both the non X-ray instrumental background (NXB) as well as the cosmic X-ray background (CXB). We used the corresponding tuned background files provided by the \suzaku team at the \textit{HEASARC}\footnote{\url{http://heasarc.gsfc.nasa.gov/docs/suzaku/analysis/pinbgd.html}} for the spectral analysis. The source is detected in the energy range 15.0-25.0 keV. There is significant discrepancy between the HXD and other instruments that may be from calibration. We make note of these in the analysis.

\subsection{\nustar}

\nustar observed \zw229 for $\sim$20 ks in February, 2018 with its two co-aligned focal plane modules FPMA and FPMB. We analyzed the dataset with the \nustar Data Analysis Software \emph{(nustardas)} package and the updated calibration files as of April 19, 2018.\newline
We generated the cleaned and calibrated photon event files with the script \texttt{NUPIPELINE}. We used the task \texttt{NUPRODUCTS} to extract the source spectra and lightcurves from the circular region of radius 40\arcs centred on the source while the background  spectra and lightcurves were derived from the same sized circular region free from the source contamination and also avoiding the detector edges. The background-subtracted FPMA and FPMB lightcurves were generated using the ftools task \texttt{lcmath}. The source is detected between 3.0-20.0 keV.

\subsection{\swift}
\swift observed \zw229 several times over nine years resulting in 77 observations with the most recent observation simultaneous with \nustar on 11th of February, 2018. Given the short length of individual exposures, we used the available observations to retrieve the long-term XRT lightcurve. We employ the web tool Swift-XRT data products generator\footnote{\url{http://www.swift.ac.uk/user_objects/}} \citep{Evans+2007, Evans+2009}, which takes additional finer steps like pile-up correction and thus outputs the calibrated (in this work, as of November 13, 2017), background-subtracted products for the list of input target ids. UVOT observations were used to prepare the optical-UV lightcurve.\newline
\indent \swift observations of \zw229 have six different target IDs. Table~\ref{tab:datalog} presents the target IDs with corresponding frequency of observations separately shown in Column 3. The individual exposure times range as short as $\sim$ $8$ s to as long as $\sim$ $8$ ks with average times of $\sim$1-2 ks. The total exposure time is $92.46$ ks and the source is detected between 0.5-7 keV. \newline
\indent The analysis also includes the Swift-BAT spectrum that was generated from the 70-month survey\footnote{\url{https://swift.gsfc.nasa.gov/results/bs70mon/}}\citep{Baumgartner+2013}. 

\subsection{Spectral modeling}
Spectral fittings have been performed using \texttt{XSPEC v12.9.1}, \cite{Arnaud+1996}. \xmm EPIC PN, \suzaku XIS and \nustar spectra were binned using the optimal binning technique described in \citet{Kaastra+2016} and implemented with the \texttt{Python} code made available by C. Ferrigno\footnote{\url{https://cms.unige.ch/isdc/ferrigno/developed-code/}}. The average \swift source spectrum was grouped such that ``C-statistic" can be applied. \newline

\indent The model likelihoods were determined using the \citet{Cash1979} statistic, which is modified in {\sc{XSPEC}} as the $C$-statistic. However, note the BAT data are Gaussian\footnote{\url{https://swift.gsfc.nasa.gov/analysis/threads/batspectrumthread.html}}, but we continue to present $C$-statistics here as well for simplicity. We also ignored bad channels from our spectral analysis. 

The Galactic column density of $5.83\times10^{20}$ cm$^{-2}$ towards \zw229 \citep{Kalberla+2005}, is kept fixed in all spectral fits. The fluxes in the relevant spectral models were calculated using the {\sc{XSPEC}} model {\sc{cflux}}. 

As the \suzaku data were taken in XIS-nominal position, the HXD-PIN and XIS cross-normalisation is fixed to 1.16. \nustar FPMA and FPMB spectra are individually fitted with the cross-normalisation constant fixed to 1 for FPMA and set free for FPMB. \newline \indent In the forthcoming sections, the references to \xmm data will correspond to the EPIC PN camera, \suzaku to the combined XIS0+XIS3 data, and \swift to the XRT data. The PIN data appear discrepant with other data sets. The PIN spectrum is consistently different from the Suzaku FI models and the non-contemporaneous \swift BAT and \nustar spectra. It is not certain if this is due to particularly low signal-to-noise, calibration, or intrinsic variability, but given the limited data we decide not to consider it further.

Best-fit parameters are reported in the rest frame of the source. The uncertainties on the model parameters are quoted at the 90 per cent confidence level using the Monte Carlo Markov Chains (MCMC) technique embedded in {\sc{XSPEC}}. MCMC method determines the errors on the model parameters simultaneously and renders better sampling of the parameter space as compared to other traditional methods. We use the affine invariant ensemble sampler \citet{Goodman+2010} to better deal with degeneracies between the model parameters. For good sampling, we  have chosen the number of walkers to be more than twice the number of free model parameters, with the total chain length at least 10000 or more in all the cases for the chains to converge on the same parameter values. The first 1000 steps in each run are discarded for the burn-in period to remove bias introduced due to choice of starting location. To ensure that the walkers have adequately sampled the parameter space, we have checked (using the xspec command chain stat on parameter) whether the steps were rejected less than 75 per cent of the time (which means less than 75 per cent of steps in the chain were simply repeats of the previous step's values).

\label{sect:observations}
\begin{table*}
	\begin{center}
		\caption{X-ray observations of \zw229 used in this work. \xmm data corresponds to the EPIC PN detector. \suzaku observations indicate the data taken by two front-illuminated (FI) detectors, XIS0 and XIS3, and the HXD-PIN detector. Both XRT and BAT data onboard \swift were used. For the \swift target IDs with multiple observations, only the first and the last observation ID are shown with intermediate observation IDs shown as dots. The number in brackets under Column (3) represents the total number of \swift observations for each Target ID. Column (8) indicates the energy ranges used for the broad band spectral fitting. } 

		\begin{tabular}{cccccccc}                
			\hline
			(1) & (2) & (3) & (4) & (5) & (6) & (7) & (8)\\
			Observatory & Observation ID &Frequency & Start Date & Duration & Exposure & Counts & Energy Band \\
			& & & (yr.mm.dd) & (s) & (s) & & (keV) \\
			\hline
			\hline
			\xmm EPIC PN & 0672530301 & & 2011.06.05 & 29117 & 22282 & 53786 & $0.3 - 10$ \\
			\hline
			\suzaku FI-XIS~(XIS0+XIS3) & 706035010 && 2011.06.03 & 397564 & 337361 & 86006 & $0.7 - 10$ \\
			
			\hline
			\swift XRT & 00090185001 &(1)  &2009.06.18 &  & 5065 & & \\ \\
			& 00039852001 &(2) & 2009.07.07 &  & 7672 & & \\
			& 00039852002& & 2009.07.10 &  & 3536 & & \\ \\
			& 00041246001& (17) & 2011.01.26 &  & 1742 & & \\ \\
			& 00091030001& (49)& 2011.05.29 &  & 1193 & & \\
		    & ... & & & & & & \\
		    & 00091030049 && 2011.07.02 &  & 1108 & & \\ \\
			& 00041246002& & 2011.07.03 &  & 1012 & & \\
			& ...& & & & & & \\
			& 00041246020& & 2011.07.22 &  & 932 & & \\ \\
                        & 00048585001 &(2)& 2012.08.19 &  & 1101 & & \\ 
			& 00048585002& & 2012.08.23 &  & 712 & & \\ \\
			& 00081228001 &(1) & 2018.02.11 &  & 7170 & & \\
			& Total && 2009$-$2018 &  & 92460 & 7000 & $0.5 - 7$ \\
			\swift BAT &&  &   $70$ months &  & & 132 & $15 - 150$ \\
			\nustar - FPMA & 60160705002 && 2018.02.11 & 44117 & 21992 & 1752 & $3 - 20$ \\
			\nustar - FPMB & 60160705002 && 2018.02.11 & 44117 & 21945 & 1811 & $3 - 20$ \\
			\hline
			\label{tab:datalog}
		\end{tabular}
	\end{center}
\end{table*}

\section{The Deep observations in 2011}

In 2011 \zw229 was observed with both \suzaku and \xmm for about $400$ and $30\ks$, respectively. The \xmm observation occurred during the long \suzaku pointing.

\subsection{Rapid X-ray Variability}
\label{sect:var}
\begin{figure*}
	\includegraphics[width=0.6\linewidth]{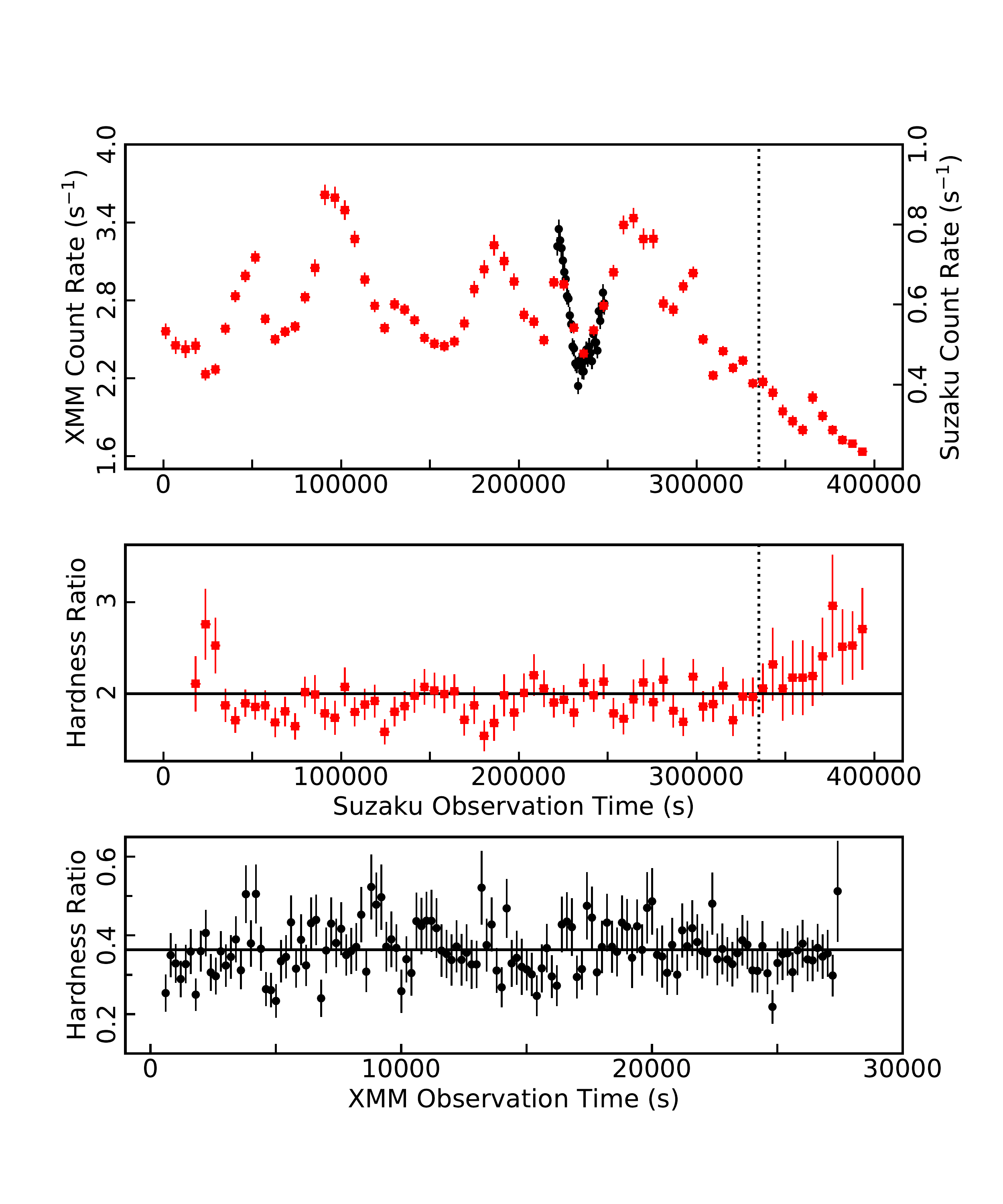}
	\caption{Upper panel: Short term X-ray light curve for \zw229. The \xmm (in black circles) count rate is depicted on the right y-axis. The binsize for the full \suzaku (red squares) light curve is set to the Suzaku orbital period of 5760 s while 200 s is selected for the \xmm light curve. The \xmm data are in agreement with the simultaneous \suzaku observation as seen in the figure. Middle panel: The hardness ratio is calculated using the 2-4 keV (hard band) and 0.5-1 keV (soft band) energy bands of the \suzaku observation. The hardness ratio is consistent with a constant ($HR$ = 2.0) for the first 335 ks of the observation while the light curve flickers by about a factor of 2 (min-to-max), before hardening in the final $\sim$ 58 ks as the count rate steadily drops. Lower panel: The hardness ratio for the XMM-Newton is fairly constant and does not show any spectral change.}
\label{lc_hr}
\end{figure*}

Figure~\ref{lc_hr} shows the background-subtracted light curves and hardness ratio ($HR=\frac{H}{S}$) variations in order to characterize its rapid variability. 
The broad band light curves exhibit persistent brightness changes by $\sim50$ per cent on time scales of hours. Despite the substantial brightness flickering, the long \suzaku $HR$ variability curves are constant during the first $335\ks$ of the observation indicating there is no significant change in the spectral shape. However, during the final $58\ks$ of the observation the AGN brightness drops off steadily and the spectrum hardens (i.e. $HR$ increases). We have checked that there is no technical issue e.g. pointing or background that can be attributed to the dip seen in the last 58 ks, and that the dip is indeed genuine.

\begin{figure}
	\includegraphics[trim=1cm 0cm 2cm 0cm,clip=true,width=0.4\textwidth]{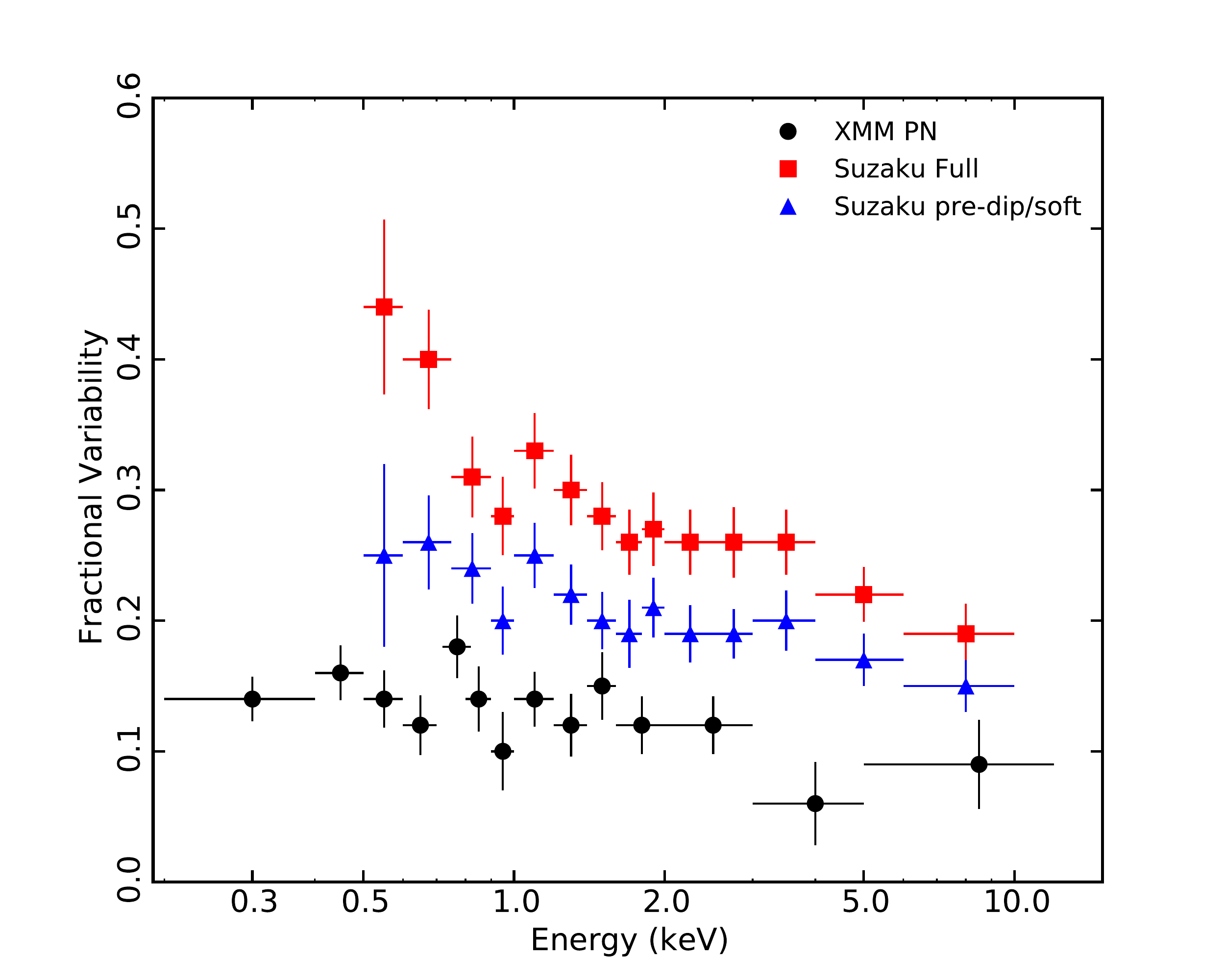}
	\includegraphics[trim=1cm 0cm 2cm 0cm,clip=true,width=0.4\textwidth]{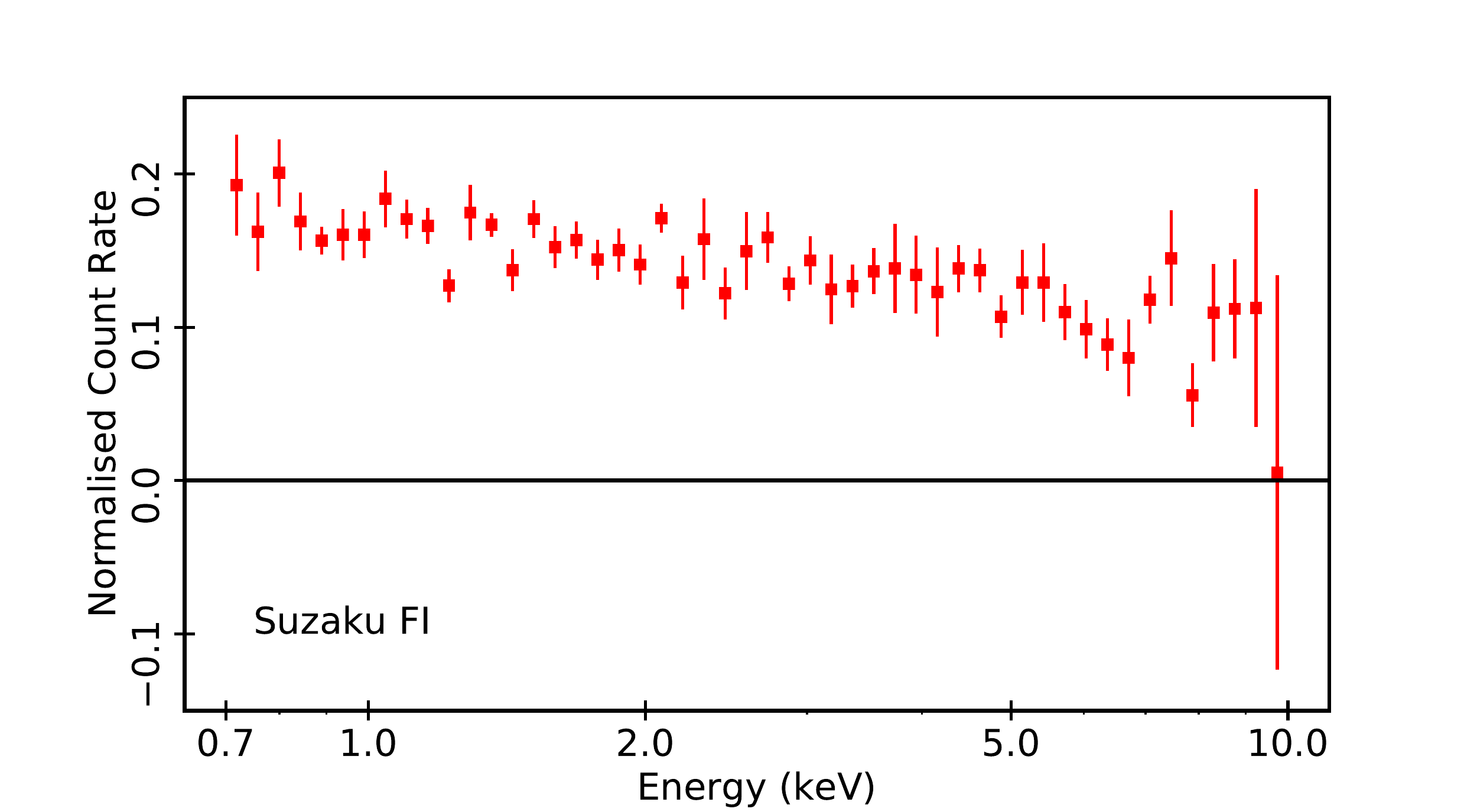}
	\caption{Upper panel: Fractional Variability for Suzaku full (red, filled squares), pre-dip (minus last 58 ks) (blue, open squares) and \xmm (black, filled circles) observations. The high variability seen in the \suzaku full data as compared to \suzaku pre-dip (soft state) data suggests that the major spectral variations stem from the last 58 ks data. Lower panel: PCA analysis of \zw229. As can be seen in the figure, the component slopes down as the energy increases. }
\label{PCA}
\label{Fvar}
\end{figure}

A fractional variability analysis ($F_{var}$) \citep{Edelson+2002}, which determines the relative strengths of observed variability across the energy bands, supports the $HR$ behaviour.   The $F_{var}$ spectrum of the \suzaku data prior to $335\ks$ show negligible spectral variations compared to the $F_{var}$ spectrum using all the \suzaku data (i.e. including the final $58\ks$) (Figure~\ref{Fvar}). The full $F_{var}$ spectrum displays larger amplitude variations at lower energies demonstrating spectral variability associated with the final $58\ks$. The \xmm data, which coincide with the first $335\ks$ of the \suzaku observation, are comparable with a constant (Upper panel; Figure~\ref{Fvar}).  

We employ Principal Component Analysis (PCA) to further investigate the spectral variability in \zw229 in a model independent manner. It is a robust technique to discern multiple spectral components existing in the spectra of a variable source \citep{Malzac+2006}. We use the PCA code based on the singular value decomposition as described in \citet{Parker+2014a} to calculate the principal components in the deep \suzaku observation. The analysis yields only one significantly variable principal component (Lower panel; Figure~\ref{PCA}), which accounts for $\approx$ 80 per cent of the variability in the spectrum.  

The primary component is greater than zero for all energies bands indicating the variability is correlated, but does exhibit a slight downward tilt from low to high energies. This is indicative of a variable component that resembles a power law, which is changing in normalisation and pivoting in a correlated manner (\citealt{Parker+2015,Gallant+2018}).

The model independent analysis of the rapid variability indicates there are two ``spectral states''. The bright high-soft state occurs in the first $335\ks$ of the observation (pre-dip in the light curve) and the dim low-hard state occurs in the final $58\ks$. Such a switch in behaviour with regards to spectral variability has been seen before in AGN, for instance in \IZw1 \citep{Gallo+2007, Gallo++2007}, \mrk530 \citep{Ehler+2018}. Moreover, the analysis suggests that the component driving the rapid variations (brightness and spectral) is the power law slope and flux.

\subsection{Characterising the X-ray spectrum}

\begin{figure}
   \centering         
  
   	\includegraphics[trim=1.0cm 0cm 1.0cm 0cm,clip=true,width=0.5\textwidth]{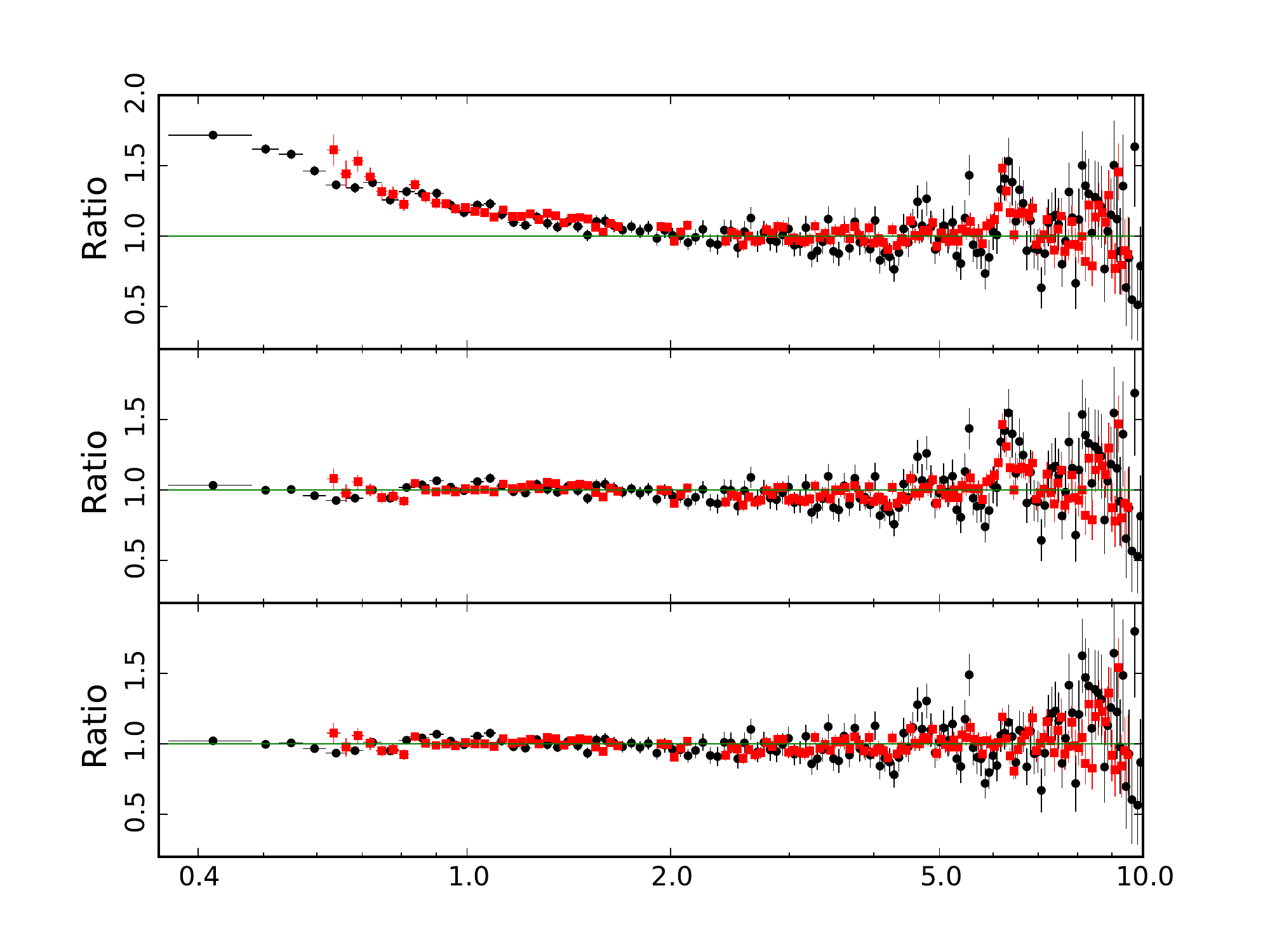}
   \caption{The ratio (data/folded model) plots for average \xmm  (black filled circles) and \suzaku  (filled red squares) spectra. Ratio plot in the top panel generated by fitting an absorbed power law model over 2-5 keV energy range suggests the presence of multiple components in the soft X-rays and the excess residuals between $\sim$ 6-7 keV. Second panel shows the ratio residuals generated by a powerlaw and black body over the usable energy ranges in \xmm and \suzaku. Lower panel shows the ratio residuals by further adding a Gaussian component at $\sim6.4\keV$. }
   \label{ratio}
\end{figure}

Fitting the average 2-5 keV \xmm and \suzaku\ spectra with a simple power law ($\Gamma\approx1.6$) modified by Galactic absorption, and extrapolating the model to lower energies appropriate for each dataset (first panel of Figure~\ref{ratio}) reveals a soft excess below 2 keV in both spectra. The addition of a blackbody component with a temperature of $kT=0.10\pm0.01\keV$ substantially improves the fit, but there are notable excess residuals around $6-7\keV$ that are likely from Fe~K$\alpha$ emission. The addition of a Gaussian profile at $E=6.49\pm0.06\keV$ and with width $\sigma=0.26^{+0.11}_{-0.09}\keV$ improves the fit further. The simple model represents the $\sim 0.3-10\keV$ spectrum well (Figure~\ref{ratio}). The corresponding C-stat is 364.24/258 d.o.f..

The smooth, extended transition from the power law to the soft excess below $\approx2\keV$ is similar to AGN like \Akn120, \mrk530 but phenomenologically different from Seyfert~1 galaxies that exhibit a sharper rise starting between $1-1.5\keV$ (for instance, \ngc4051, \cite{Pounds+2004}).

\subsection{Hardness-resolved spectral analysis}
\label{XraySpec}

\begin{figure}
	\includegraphics[trim=1.2cm 0cm 2cm 0cm,clip=false,width=0.5\textwidth]{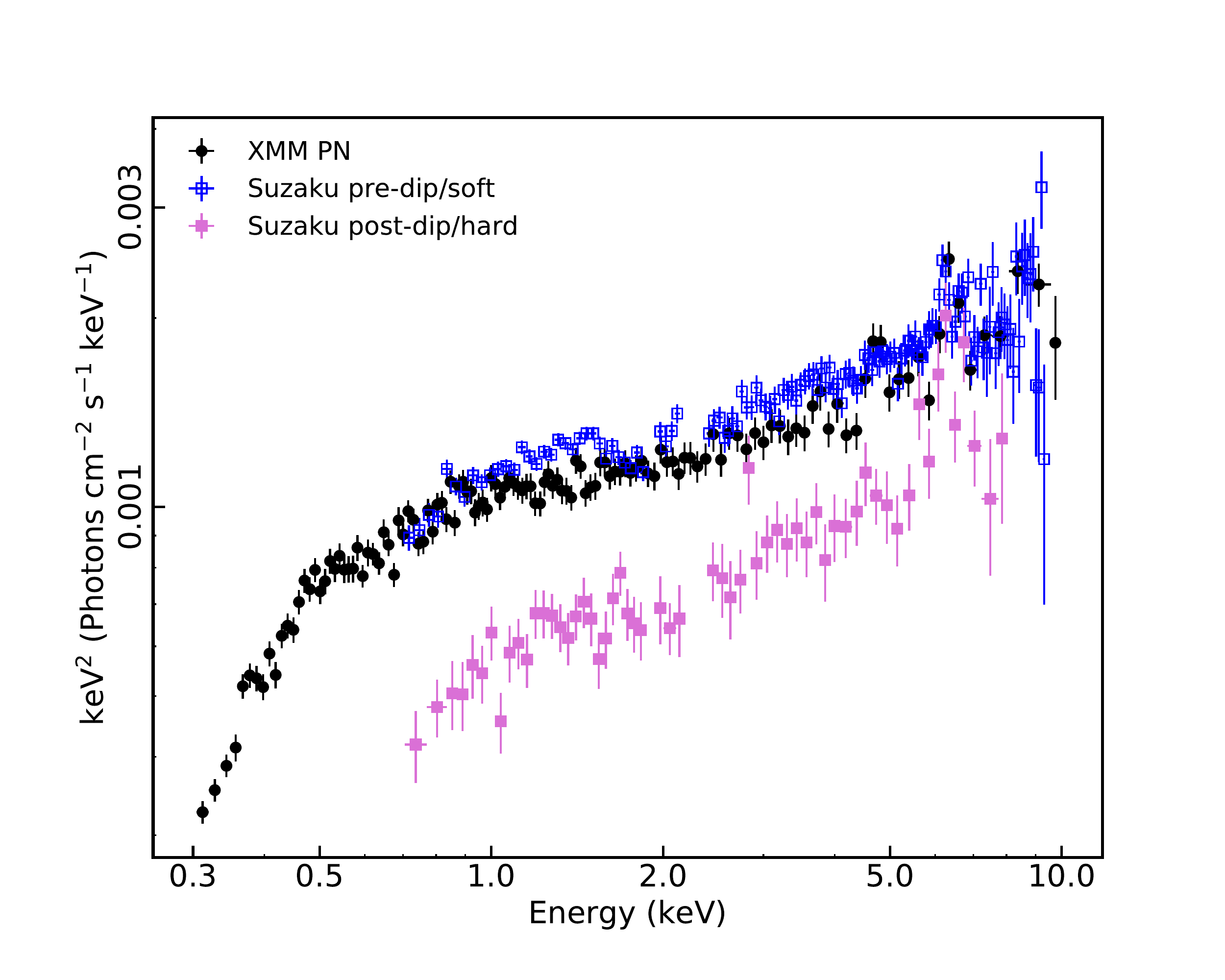}
	\caption{The unfolded spectra of \zw229 for \xmm (black circles), \suzaku pre-dip/soft (open, blue squares) and \suzaku post-dip/hard (filled purple squares) data fit by a power law with slope $\Gamma=0$ to investigate the change in the the spectral shape amongst soft and hard states. The spectra have been rebinned to ensure that trends can be seen clearly. The \xmm and \suzaku pre-dip spectra are alike and agree well with each other whereas the \suzaku post-dip data deviate significantly in nature.}
\label{eeuf}
\end{figure}

Motivated by the rapid X-ray variability, the \suzaku observation is divided into a high-soft (``pre-dip'', first 335 ks) and a low-hard spectrum (final 58 ks). The \xmm observation is entirely consistent with the \suzaku high-soft spectrum (Figure~\ref{eeuf}). In the following analysis we have thus linked the data from the two instruments allowing only a cross-normalisation constant to vary between them.

To understand the spectral appearance and behaviour of \zw229, we attempt fitting the spectra of the deep X-ray observations obtained by \xmm and \suzaku with physical models in a self-consistent manner. Specifically, we examine three physical scenarios to describe \zw229: partial covering absorption (e.g. \citealt{Tanaka+2004}), blurred reflection (e.g. \citealt{Ross+2005}) and soft Comptonisation (e.g. \citealt{Done+2012}). Each model provides a different interpretation of the soft excess thus may reveal the physical origin of the emission below $2\keV$. Given the presence of a relatively narrow $\sim6.4\keV$ emission feature, each model includes a neutral and unblurred reflection component (i.e. {\sc{xillver}} in {\sc{XSPEC}}; \citealt{Garcia+2011, Garcia+2013}) to account for emission from the distant torus. The torus is assumed to be cold ($\xi = 1 \ergcmps$, $\xi=4\pi F/n$ where $\xi$ is the ionization parameter, n is the hydrogen number density and F is the incident flux) with solar abundances of iron ($A_{Fe}=1$).

None of the three models constrain the line-of-sight inclination well. In models that the inclination is a parameter, it is fixed at $45\deg$. Similarly, models (i.e. soft Comptonisation and blurred reflection) that measure the black hole spin parameter `a' ($a\equiv Jc/GM^2$ where J and M are the angular momentum and mass of the black hole respectively) do not constrain this parameter well. The limited contemporaneous UV data (i.e. only UVW1 and UVW2) makes it difficult for the soft Comptonisation models ({\sc{optxagnf}}) to constrain spin. OM data are not used in the fit. Likewise, the weak Fe~K$\alpha$ feature makes it challenging to measure spin with the blurred reflection models (e.g. \citealt{Bonson+2016}). In these cases, three fits are attempted with the spin parameter frozen at a=0.0, 0.5 and 0.998, and the best fit is reported.

\subsubsection{Ionized Partial Covering Absorption}
\label{absorption}

In the ionized partial covering scenario (e.g. \citealt{Tanaka+2004}; \citet{Reeves+2008}; \cite{Miller2007, Miller+2009, Gallo++2011, Gallo+2011}; \citealt{Gallo+2004}), the power law dominated continuum is partially obscured some fraction f$_{cov}$ by an ionised medium. The observed X-ray spectrum is then a combination of obscured emission and direct power law emission.

The soft and hard spectra can be fitted well with a consistent partial covering model. The model used is {\sc{tbabs*zxipcf*(cutoffpl + xillver)}}. The best fit requires only one absorber which is highly ionized ($\xi=16.0^{+1.6}_{-1.3}\ergcmps$) and covers $f_{cov}=0.37\pm0.07$ of the power law source. Adding a second partial covering absorber does not significantly improve the fit further.

\begin{figure}
   \centering         
  
   	\includegraphics[trim=3cm 0cm 2cm 0cm,clip=true,width=0.5\textwidth]{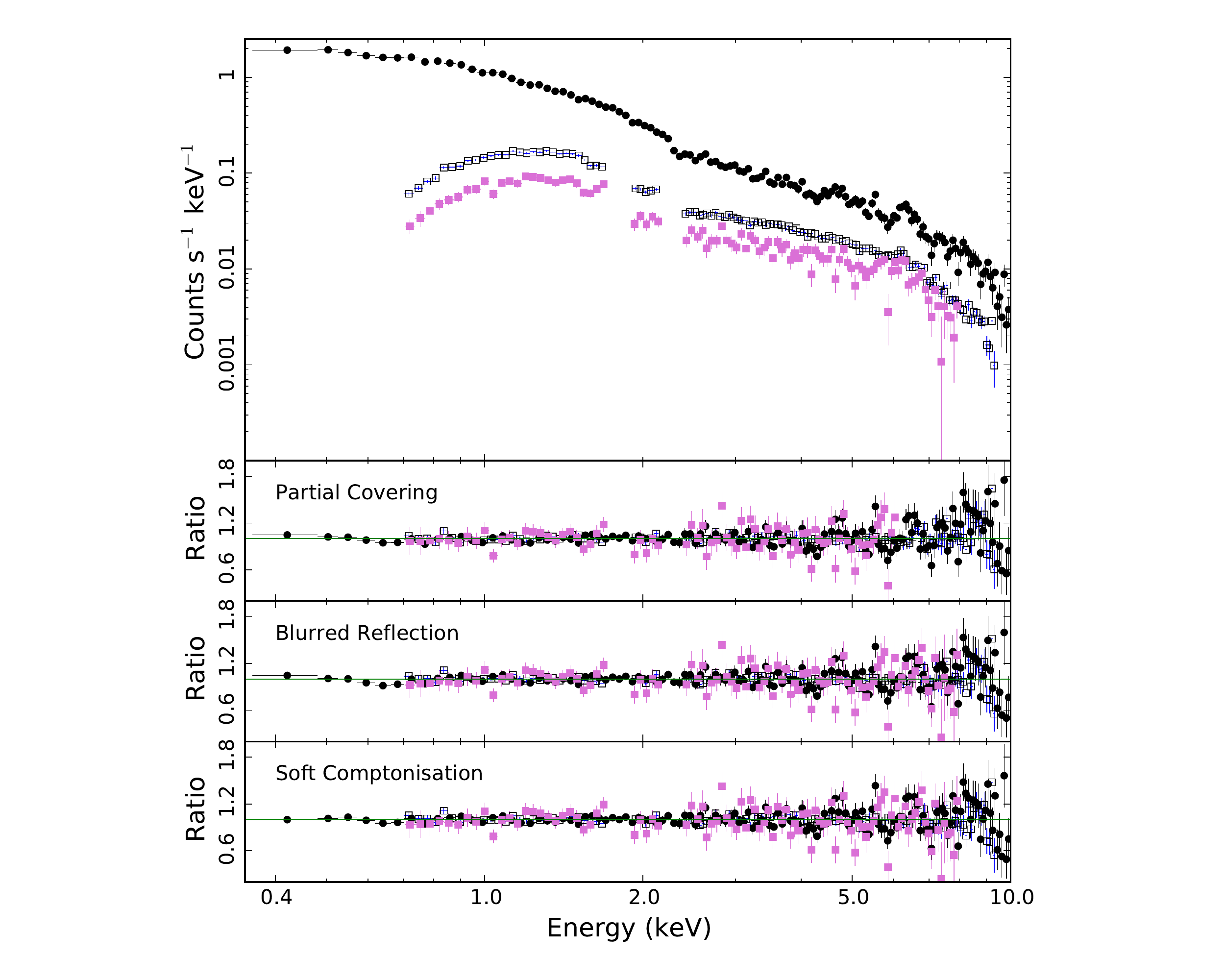}
   \caption{The folded spectrum (top panel) and ratio plot for \xmm dataset (black filled circles), \suzaku soft (open blue squares) and \suzaku hard (filled purple squares). The bottom three panels show the ratio from fitting the spectra with the partial covering, blurred reflection, and soft Comptonisation models described in Section 3.} 
   \label{foldspec-PC}
\end{figure}

Multiple scenarios are considered to describe the changes from the soft to hard states in the AGN. This can be achieved by changing the values of the absorber (e.g. covering fraction and column density). However, a better fit can be reached by allowing the photon index and normalisation of the intrinsic power law to change instead (Figure~\ref{foldspec-PC} first and second panels; Table~\ref{PCtab:fits}). This scenario is also in agreement with the timing analysis in Section~\ref{sect:var}.
\begin{table*}
\caption{The best-fit model parameters for the ionized partial covering, blurred reflection and soft Comptonisation of \zw229 for deep \xmm and \suzaku observations.  
The model, model components and model parameters are listed in Columns 1, 2 and 3. 
The subsequent columns refer to \xmms, \suzaku soft and \suzaku hard datasets respectively.  
The values of parameters that are linked between datasets appear in only one column.  
The fixed parameters are denoted by superscript $f$. The normalisation of the power law component is photons~keV$^{-1}$~cm$^{-2}$~s$^{-1}$ at 1 keV.
}
\centering
\scalebox{1.0}{
\begin{tabular}{cccccc}                
\hline
(1) & (2) & (3) & (4) & (5) & (6) \\
Model & Model Component &  Model Parameter  &  XMM & Suzaku-soft & Suzaku-hard \\
\hline
Single Ionised & Power law & $\Gamma$ & $1.95^{+0.03}_{-0.02}$ &  & $1.81^{+0.06}_{-0.05}$ \\
partial covering & &$Norm~(10^{-3})$& $1.90\pm0.09$ & & $0.93^{+0.06}_{-0.05}$ \\
 & & & & &\\
\hline
 & Absorber 1& $\nh$~($\times~10^{22}~cm^{-2})$  & $58.2^{+32.1}_{-15.5}$ &  &      \\
 &           & $C_f$    & $0.37\pm0.07$ &  &  \\
 &           & $log\xi~(erg~cm~s^{-1})$   & $2.21^{+0.19}_{-0.12}$ & &    \\
\hline
 & Xillver &    $Norm~(10^{-5})$ & $1.96^{+0.52}_{-0.38}$ & & \\
 &            &  $Inc~(\deg)$ & $45^{f}$ & & \\
 &              & $A_{Fe}~(solar)$     & $1.0^{f}$ & & \\
 &              & $log\xi~(erg~cm~s^{-1})$   & $0.0^{f}$ & & \\
 &              & $E_{cut}~(keV)$   & $300.0^{f}$ & & \\
\hline
 &             Fit Quality & $C-stat$ & $454/354$ &   & \\
\hline
\hline
Relativistic reflection & Cut-off Power law & $\Gamma$ & $1.88\pm0.01$ & & $1.70^{+0.07}_{-0.08}$ \\
&             & $High~E_{cut}~(keV)$  & $300.0^{f}$ & & \\
         &  &$Norm~(10^{-4})$ & $9.60^{+0.37}_{-0.40}$ &  & $3.63^{+0.41}_{-0.42}$ \\
\hline
 & Relxill  & $log\xi~(erg~cm~s^{-1})$    & $2.56\pm0.15$ & &   \\
&            & $Norm~(10^{-5})$     & $2.60^{+0.37}_{-0.23}$ & &   \\
 &           & $R_{in}$ ($isco$)   & $-1.0^{f}$               &                        &               \\
 &           & $R_{out}$ ($\rg$)   & $400^{f}$                &                        &               \\
 &           & $index1$   & $6^{f}$               &                        &                \\
&           & $index2$   & $3^{f}$               &                        &                \\
&           & $R_{br}$ ($\rg$)  & $10^{f}$               &                        &                \\
&           & $a$   & $0.998^{f}$               &                        &                \\
\hline
 & Xillver &    $Norm~(10^{-5})$ & $1.70^{+0.31}_{-0.33}$ & & \\
 &           &   $Inc~(\deg)$ & $45.0^{f}$ & & \\
\hline
 & Reflection Fraction &    $(0.1-100)~keV$ & $0.4\pm0.05$ &$0.5\pm0.09$ & $0.3\pm0.2$\\
\hline
 &             Fit Quality & $C-stat$ & $411/355$ &   & \\
\hline
\hline
Comptonisation & Hard X-ray continuum & $\Gamma$ & $1.71\pm0.04$ & & $1.51\pm0.02$ \\
 (Low spin) & &$f_{pl}$ & $0.84^{+0.01}_{-0.02}$ &  & $0.89\pm0.03$  \\ 
&             & $r_{cor}$ ($\rg$)  & $11.10^{+0.50}_{-0.41}$ \\        
\hline
 & Soft excess  & $kT_{e}~(keV)$   & $0.59^{+0.08}_{-0.07}$ &    \\
 &           & $\tau$   & $8.45^{+0.73}_{-0.70}$  &              \\

\hline
 & Xillver &$Norm~(10^{-5})$ & $1.25^{+0.24}_{-0.26}$ &  \\
 &           &   $Inc~(\deg)$ & $45.0^{f}$ &  \\
&             & $A_{Fe}~(solar)$    & $1.0^{f}$  & \\
&             & $E_{cut}~(keV)$   & $100.0^{f}$  & \\
\hline
 & Other parameters   & $M_{BH}$ ($M_{\odot}$)   & $10^{7}$     \\
&           & $logr_{out}$ ($\rg$)   & $5^{f}$                           \\
&             & $logL/L_{Edd}$  & $-0.90^{f}$ \\  
&             & $a$  & $0.5^{f}$ \\ 
&             & $dist$ ($\Mpc$)  & $121.2^{f}$ \\ 
\hline

&             Fit Quality & $C-stat$ & $395/354 $ &   & \\
\hline

\label{PCtab:fits}
\end{tabular}
}
\end{table*}

\subsubsection{Relativistic Reflection}
\label{reflection}

As emission from the primary power law component (e.g. corona) illuminates the colder accretion disc a reflection spectrum will be produced (e.g. \citealt{Ross+2005}). The backscattered emission will be blurred because of motions in the accretion disc close to the black hole. This could generate a smooth spectrum that reproduces the soft excess emission. 

To investigate the  presence of relativistic reflection in \zw229 spectra, we adopted a spectral model comprising of two independent components to describe the continuum. A powerlaw ({\sc{cutoffpl}}) illuminating the disk and a component for relativistically broadened reflection accounted by {\sc{relxill}} \citep{Dauser+2010, Dauser+2013}. Since emission from a relativistically blurred Fe~K$\alpha$ line is not significant in the spectrum of \zw229, the smooth soft excess will be driving the fit.

The model {\sc{relxill}} assumes a power-law form of the reflection emissivity from the accretion disc $E(r)\propto r^{-q}$ where q is the emissivity index and E is the emissivity of the gas due to reflection. In the general relativistic regime, the inner emissivity index steepens up to breaking radius R$_{br}$ and then tends to take constant index of {value $\approx$ 3 consistent with the Newtonian regime \citep{Wilkins+2011}. We have fixed values for q1 and q2 consistent with this scenario for spins a=0.0, 0.5 and 0.998. 

The best statistical fit is achieved for blurred reflection from a highly ionised disc ($\xi=363\pm1\ergcmps$) around a maximum spinning black hole ($a=0.998$). The quality of the fit diminishes when smaller values of the spin parameter are attempted ($\Delta~C=57,46 $ for spin a=0.0 and 0.5, respectively for the same number of degrees of freedom). 

The difference between the soft and hard spectrum can be explained by changing the ionisation parameter and normalisation of the reflection component, but again a much better fit is achieved when the power law component changes in flux and photon index ($\Delta~C=23$ for same number of degrees of freedom; Figure~\ref{foldspec-PC} (third panel); Table~\ref{PCtab:fits}). This is rather curious as one expects the ionisation of the disc to respond to changes in the fluctuating power law as has been seen in other AGN (e.g. \citealt{Bonson+2018, Chiang+2015}).

\subsubsection{Soft Comptonisation}
\label{Comptonisation}
As described in previous sections, it is evident that both the partial-covering and the relativistic reflection scenarios seem to provide good statistical fits to the soft and hard states of \zw229. 

Another context that needs to be addressed in order to explain the nature of the X-ray emission is the process of intrinsic disc Comptonisation. Using {\sc{optxagnf}} in {\sc{XSPEC}} \citep{Done+2012}, this model characterizes the physical nature of broad-band emission from three components: 1) accretion disc 2) optically thick, low temperature thermal Comptonisation that produces the soft X-ray excess and 3) an optically thin, high temperature thermal Comptonisation to produce the hard X-ray continuum. In this model, the flux (normalisation) is set by the physical parameters: black hole mass ($M_{BH}=10^{7} M_{\odot}$), Eddington ratio ($L/L_{Edd}=0.123$, \citealt{Smith+2018, Buisson+2017}), black hole spin ($a$) and luminosity distance ($D_{L}=121.2 \Mpc $). 

The model is examined at three different spin parameters. The best-fit is achieved at an intermediate spin value (a=0.5) though it is not significantly better than for zero spin  ($\Delta~C=0.3$) or maximum spin values ($\Delta~C=6.90$).  In the case of maximum spin, we are unable to constrain coronal radius parameter ($r_{cor}$) as it pegs at the highest limit. This might be a limitation in the model rather than for physical reasons (e.g. \citealt{Thomas+2016, Collinson+2017}). Changes in the normalisation and photon index of the power law are sufficient to describe the differences between the soft and hard states in \zw229 (C-stat=395/354 d.o.f.; Figure~\ref{foldspec-PC} (fourth panel); Table~\ref{PCtab:fits}). 

Based on examination of the deep 2011 observations, several physical models could describe the X-ray spectrum. Statistically, the best fit is from the soft-Comptonisation model where the soft-excess arises from a cooler Comptonising layer above the accretion disc. In all models, the variability is best described by changes in the brightness and spectral shape ($\Gamma$) of the intrinsic power law.

\section{Multi-epoch observations between 2009-2018}
\label{var}

The rapid variability appears dominated by changes in the intrinsic power law independent of the spectral model adopted to fit the data. In this section, we examine the long-term spectral variability between 2009 and 2018.

\subsection{Long Term, broadband Variability}
To examine the long term variability behaviour, we take into account the 9 year long observations of \zw229 by \swift XRT, together with the deep observations of the source with \suzs, \xmm and \nustar. The 0.3-10 keV count rates of \zw229 spanning nearly 9 years is presented in Figure~\ref{fluxall} (top panel). The X-ray lightcurve exhibits significant variability over yearly time scales, but shows an extended period of approximately 30 days in 2011 where the count rate drops to about one-tenth the average (Figure~\ref{fluxall}). 

The variability is also notable in the optical-to-UV bands (Figure~\ref{fluxall} middle panel). The simultaneous X-ray and UV measurement allow us to gauge the broadband spectral variability. For this, the hypothetical power law between 2500 \A and 2 keV (i.e. $\alpha_{ox}$, \citealt{Tananbaum+1979}) is estimated from the simultaneous X-ray and UV observations. The measured $\alpha_{ox}$ is steeper during the low-flux state in 2011 indicating the X-rays have dimmed more than the UV. 

Comparing our measurements of $\alpha_{ox}$ to the expected values given the UV luminosity of the AGN (e.g. \citealt{Strateva+2005}), provides us with a measure of $\Delta\alpha_{ox}$ $(\Delta\alpha_{ox} = \alpha_{ox} - \alpha_{ox}(L_{2500\A}))$.  This is an indicator of the X-ray strength compared to the UV (Figure~\ref{fluxall} lower panel). \zw229 displays values of $\Delta\alpha_{ox}\approx0$ during the \suzs, \xmms, and \nustar epochs indicating that it appears like a typical or ``normal'' AGN.

During the low X-ray flux state in 2011, $\Delta\alpha_{ox}$ drops to $\sim-0.11$, indicating the AGN has become X-ray weak. According to \citet{Gallo2006}, X-ray weakness is often accompanied by X-ray spectral complexity that can be attributed to enhanced blurred reflection or absorption. Similar long term variability studies of various sources dropping to X-ray weak states (with respect to the AGN's UV flux) and low flux states have been carried out in the past \citep{Gallo+2011, Gallo++2011, Grupe+2012, Miniutti+2009, Schartel+2010, Gallo+2015}.

Unfortunately, the X-ray weak state in \zw229 was only observed with snap-shot \swift observations.  It is not possible to extract a significant spectrum of the X-ray weak state. 

\begin{figure*}
	\includegraphics[width=0.8\linewidth]{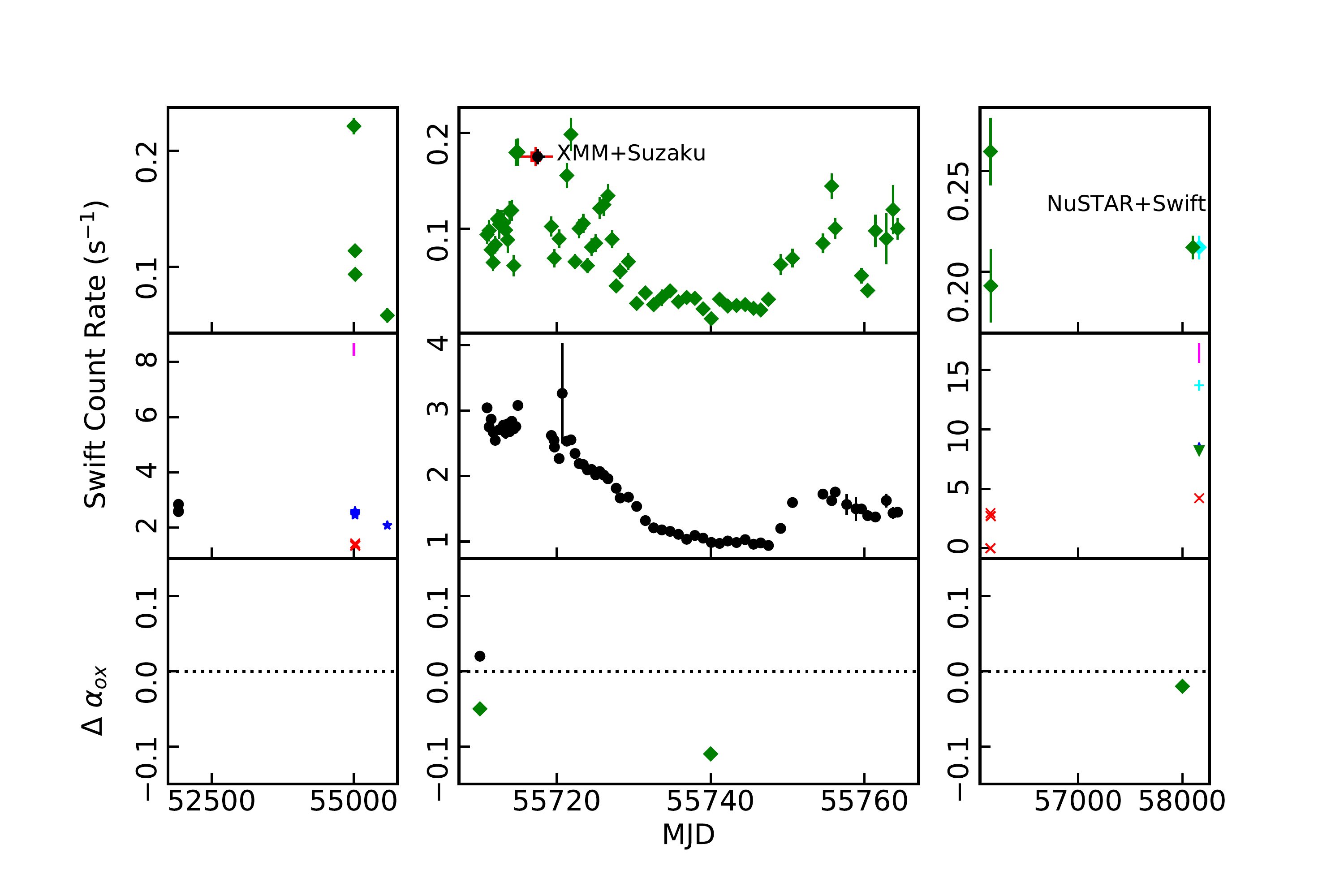}
	\caption{Top panel: The 9-year long term light curve for \zw229. The average 0.3-10 keV count rates for \xmm, \suzaku and \nustar are normalised to \swift XRT count rate. The X-axis represents continuous time of observations in MJD but scales differently in each panel for the sake of clarity (MJD: Modified Julian Date = JD - 2,400,000.5). The source shows substantial variability with sporadic transitions to-and-from high and low X-ray flux states. Middle panel: The long term ultraviolet count rate for all the \swift UVOT filters (black `circles' for UVW2, red `crosses' for UVM2, blue `stars' for UVW1, magenta `vertical lines' for U, cyan `pluses' for B and dark green `triangles' for V filter) within the respective MJD times. Lower panel: The plot of $\Delta\alpha_{ox}$ with time. Horizontal dashed line at $\Delta\alpha_{ox}=0$ represents a `normal' AGN given the UV luminosity of \zw229 (e.g. \citealt{Strateva+2005}). Negative values indicate the AGN is X-ray weak.  }
\label{fluxall}
\end{figure*}


\subsection{Multi-epoch Spectral modelling}
\label{multiepochspec}
Multi-epoch modelling is extremely valuable to constrain the physical parameters that vary on different timescales. One can link the spectral parameters that do not vary over the long timescales, for instance, black hole spin, disc inclination and elemental abundances thus simplifying the spectral fitting procedure in an otherwise complex model.

We jointly fit the \xmms, the \suzaku high-soft and the \suzaku low-hard spectra alongwith the \nustar FPMA and FPMB. The model parameters of the \nustar FPMB power-law parameters were linked to that of FPMA. However, to show the best-fit results, we do not consider the average \swift XRT and BAT in the multi-epoch spectral analysis simply to avoid mixing of different flux states during 9 years of \swift observations. BAT spectrum is consistent with the multi-instrument spectra as shown in Figure~\ref{foldspec-all-pc}. The \xmm and \suzaku cross-normalisation is treated as in Section~\ref{XraySpec}. 
Each model includes a neutral and unblurred reflection component (i.e. {\sc{xillver}} to account for distant reflection. As before, the torus is assumed to be cold ($\xi = 1 \ergcmps$) with solar abundances of iron ($A_{Fe}=1$), inclination fixed to $45\deg$. Since \nustar is not sensitive to energies below $\sim$ 3 keV, the value of the power law fraction that is from the high-temperature component ($f_{pl}$) pegs at the upper limit (1.0). 

The analysis reveals similar results as that of the deep observations. Despite the inclusion of spectral data up to $25\keV$, the three physical models are not constrained significantly better. Between the epochs, we attempt to vary the inner disc ionization parameter, the absorber covering fraction and distant reflection in addition to varying the power-law parameters. Soft-Comptonisation and blurred reflection models describe the long-term observations and variability best (Figure~\ref{foldspec-all-pc}, Table~\ref{allpctab:fits}).  Moreover, like the rapid variability, the long-term spectral variability over years appears to be driven by changes in the slope and flux of the intrinsic power law component.  

\begin{landscape}
\begin{table}
\caption{The best-fit model parameters for the ionized partial covering, blurred reflection and soft Comptonisation of \zw229 for muti-epoch observations.  
The model, model components and model parameters are listed in Columns 1, 2 and 3. 
The subsequent columns refer to \xmms, \suzaku soft, \suzaku hard, \nustar FPMA and \nustar FPMB datasets respectively.  
The values of parameters that are linked between datasets appear in only one column.  
The fixed parameters are denoted by superscript $f$. Pegged parameter is denoted by superscript $p$. The normalisation of the power law component is photons~keV$^{-1}$~cm$^{-2}$~s$^{-1}$ at 1 keV.
}
\centering
\scalebox{1.0}{
\begin{tabular}{cccccccccc}                
\hline
(1) & (2) & (3) & (4) & (5) & (6) & (7) & (8)  \\
Model & Model Component &  Model Parameter  &  XMM & Suzaku-soft & Suzaku-hard  &FPMA & FPMB\\
\hline
Single Ionised & Power law & $\Gamma$ & $1.95\pm0.02$ &  & $1.82\pm0.05$ & $2.02\pm0.08$ \\
partial covering & &$Norm~(10^{-3})$& $1.90^{+0.10}_{-0.09}$ & & $0.94^{+0.06}_{-0.05}$ &  $2.67^{+0.41}_{-0.30}$\\
\hline
 & Absorber 1& $\nh$~($\times~10^{22}~cm^{-2})$  & $55.4^{+0.80}_{-1.05}$ &  &      \\
 &           & $C_f$    & $0.37^{+0.04}_{-0.03}$ &  &  \\
 &           & $log\xi~(erg~cm~s^{-1})$    & $2.20^{+0.31}_{-0.57}$ & &    \\
\hline
 & Xillver &    $Norm~(10^{-5})$ & $2.32\pm0.34$ & & \\
 \hline
&              & Unabsorbed Flux $(10^{-12}~erg~cm^{-2}~s^{-1})$  & $8.0\pm0.1$ & &$4.7^{+0.2}_{-0.1}$ &$7.1^{+0.3}_{-0.1}$ \\
&              & Luminosity $(10^{43}~erg~s^{-1})$  & $1.41\pm0.02$ & &$0.83^{+0.04}_{-0.02}$&$1.25^{+0.05}_{-0.02}$  \\
\hline
 &             Fit Quality & $C-stat$ & $600/458$ &   & \\
\hline
 \hline
Relativistic reflection & Cut-off Power law & $\Gamma$ & $1.88^{+0.02}_{-0.01}$ & & $1.70^{+0.07}_{-0.08}$ &$1.93^{+0.11}_{-0.08}$\\
&             & $High~E_{cut}~(keV)$   & $300.0^{f}$ & & \\
         &  &$Norm~(10^{-4})$ & $9.58^{+0.21}_{-1.44}$ &  & $3.61^{+0.18}_{-1.11}$ &$13.9^{+3.25}_{-1.93}$&\\
\hline
 & Relxill  & $log\xi~(erg~cm~s^{-1})$     & $2.64^{+0.07}_{-0.17}$ & &   \\
&            & $Norm~(10^{-5})$     & $2.60^{+0.33}_{-0.28}$ & &   \\
 &           & $R_{in}$ ($isco$)   & $1.24^{f}$               &                        &               \\
 &           & $R_{out}$ ($\rg$)   & $400^{f}$                &                        &               \\
 &           & $index1$   & $6^{f}$               &                        &                \\
&           & $index2$   & $3^{f}$               &                        &                \\
&           & $R_{br}$ ($\rg$)  & $10^{f}$               &                        &                \\
&           & $a$   & $0.998^{f}$               &                        &                \\
\hline
 & Xillver &    $Norm~(10^{-5})$ & $1.78^{f}$ & & \\
\hline
&              & Unabsorbed Flux $(10^{-12}~erg~cm^{-2}~s^{-1})$  & $8.0\pm0.1$ & &$4.7\pm0.1$ &$7.1\pm0.2$ \\
&              & Luminosity $(10^{43}~erg~s^{-1})$  & $1.41\pm0.02$ & &$0.83\pm0.02$&$1.25\pm0.04$  \\
\hline
 &             Fit Quality & $C-stat$ & $561/459$ &   & \\
\hline
\hline
Comptonisation & Hard X-ray continuum & $\Gamma$ & $1.76\pm0.02$ & & $1.53\pm0.01$ & $1.82^{+0.07}_{-0.04}$\\
 (Low spin) & &$f_{pl}$ & $0.84\pm0.02$ &  & $0.89^{+0.03}_{-0.04}$ & $0.94^{p}$  \\  
&             & $r_{cor}$ ($\rg$)  & $10.81^{+0.23}_{-0.14}$ \\       
\hline
 & Soft excess  & $kT_{e}~(keV)$   & $0.49^{+0.07}_{-0.04}$ &    \\
 &           & $\tau$   & $9.26^{+1.22}_{-1.50}$  &              \\

\hline
 & Xillver &$Norm~(10^{-5})$ & $1.33^{+0.23}_{-0.15}$ &  \\
\hline
 & Other parameters   & $M_{BH}$ ($M_{\odot}$)   & $10^{7}$     \\
&             & $a$  & $0.5^{f}$ \\ 
\hline
&              & Unabsorbed Flux $(10^{-12}~erg~cm^{-2}~s^{-1})$  & $8.2\pm0.1$ & &$4.8\pm0.2$ &$7.0\pm0.2$ \\
&              & Luminosity $(10^{43}~erg~s^{-1})$  & $1.44\pm0.02$ & &$0.84\pm0.04$&$1.23\pm0.04$  \\
\hline

&             Fit Quality & $C-stat$ & $546/458 $ &   & \\
\hline
\hline
\label{allpctab:fits}
\end{tabular}
}
\end{table}
\end{landscape}
  
\begin{figure}
\centering         
\includegraphics[trim=3cm 0cm 2cm 0cm,clip=true,width=0.5\textwidth]{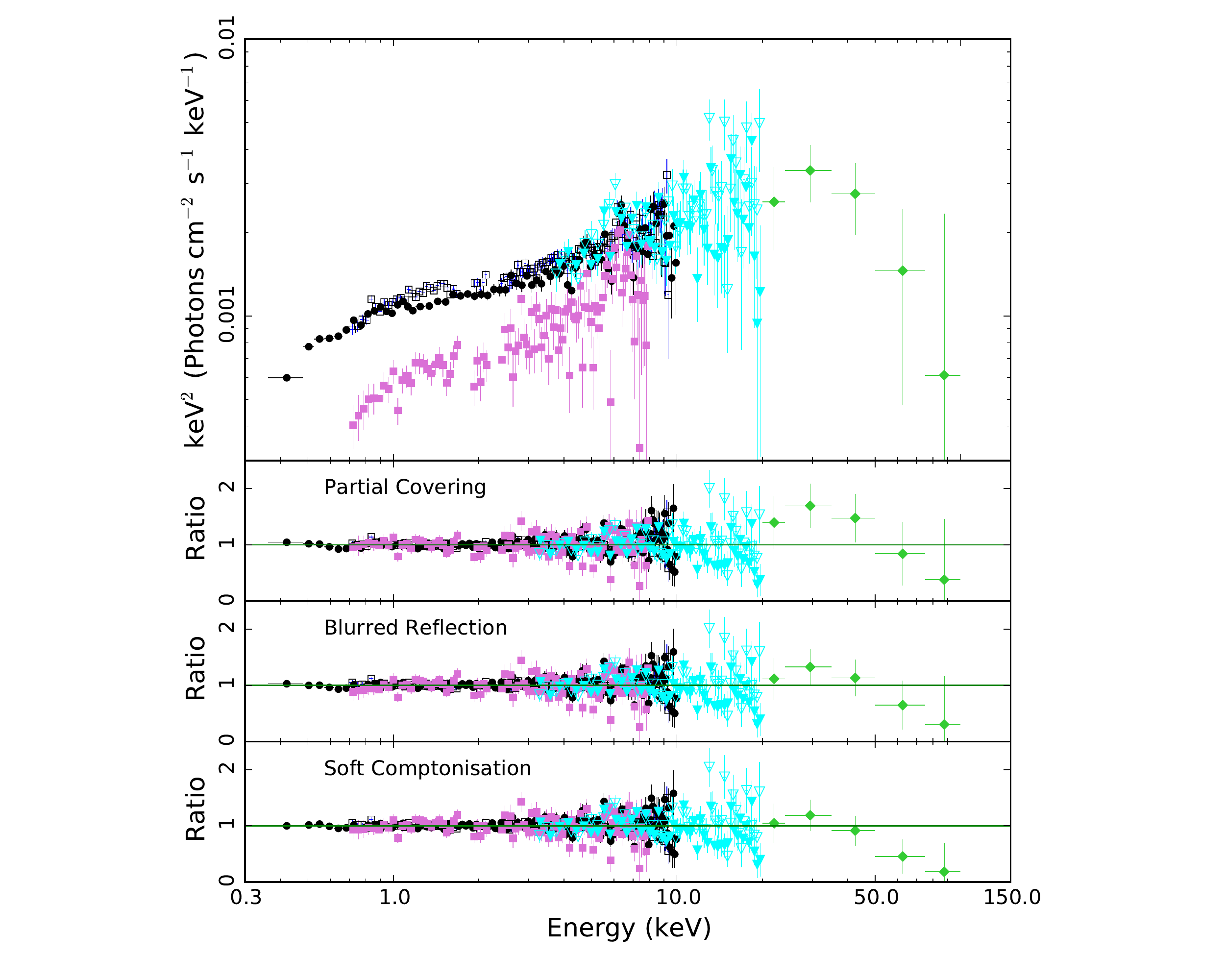}
\caption{The unfolded spectra (compared to $\Gamma=0$) showing the data from all the instruments taken between 2009-2018. Shown in the figure are \xmm (black filled circles), \suzaku soft (open blue squares), \suzaku hard (filled purple squares), \nustar FPMA (cyan, filled triangles), FPMB (cyan, open triangles) and \swift BAT (green, filled diamonds) spectra. The ratio plots are from fitting the models discussed in Section~\ref{multiepochspec} to the multi-epoch observations (see Table 3).}
\label{foldspec-all-pc}
\end{figure}


\section{Discussion}
\label{disc}
\zw229 is a Seyfert 1 galaxy with a surplus of interesting X-ray and optical/UV properties. Its astonishing optical variability behaviour has already been established by previous studies, with the clear measurement of characteristic break in the power spectrum \citep{Edelson+2014}.

\subsection{Origin of the Soft excess}

\zw229 exhibits a blend of curious features similar to features exhibited by sources like \Akn120 (e.g. \citealt{Vaughan+2004, Matt+2014, Porquet+2018}) and \mrk530 (\citealt{Ehler+2018}). \zw229 shows a soft excess that is gently increasing above the power law starting at about 2 keV toward lower energies. The broad component of Fe~K$\alpha$ line is insignificant and the strength of the Compton hump is negligible.

Spectral modelling is often challenging in such objects as there are few strong and distinct features. Indeed, the best statistical fit is obtained with the soft-Comptonisation model that, by nature, generates the smoothest spectrum.

The spectral change can also be well understood in the context of intrinsic disc Comptonisation model (Section ~\ref{Comptonisation}) where the variations in the hard X-ray power law continuum for a non-rotating black hole configuration are required to explain the observed variability in the high-soft and low-hard spectra. The value of $R_{cor}\sim11$ $r{_g}$ puts a limit to the region up to which the thermal emission can be produced. Below this radius, around 80-90 percent of the remaining energy will dissipate in the optically thin, high temperature (100 keV) corona with photon index $\sim$ 1.7 while the rest would contribute to the soft excess. The coronal radius evaluated with our analysis is consistent with the coronal size (within 20$\rg$) estimated by the time lag studies on this source \citep{Adegoke+2017}. Similarly, in the blurred reflection model, the disc is substantially ionised and the reflection fraction is rather low to account for the weak broad line and Compton hump.  Alternatively, if the blurred emission is exposed to additional Comptonisation as it emerges from the disc, this could describe the weak reflection features (e.g. \citealt{Wilkins+2015}).
\subsection{Variability from hours to years}
Analysis of the deep, long, single-epoch \suzaku observation shows for the most part the source varies rapidly in brightness, but exhibits negligible spectral variability. It is only during the final 58 ks of the observation that the AGN appears to steadily diminish in brightness and become spectrally hard. With regards to the physical mechanism causing the shift in hardness during the last 58 ks of \suzaku observation, this could possibly be due to non-uniform temperature or optical depth of the primary corona which give rise to two distinct but coordinated varying power law components. The inhomogeneous heating of the corona could arise for instance, due to different magnetic field strengths at different radii \citep{Lohfink+2016} or perhaps in a multi-component corona (e.g. \citealt{Gallo++2007, Wilkins+2017}) or possibly due to two component Comptonization scenario as outlined in \cite{Done+2012} . Examination of the hardness ratios, PCA, F$_{var}$, and spectral modelling all indicate that the variability arises from the intrinsic power law component that is changing in brightness and shape in a correlated manner (i.e. softer when brighter).

The long term \swift XRT lightcurve shows significant variable behaviour encompassed by the \xmms, \suzaku and \nustar observations. Considering its X-ray-to-UV spectral energy distribution, \zw229 displays the characteristics of a `normal' bright AGN during the contemporaneous \xmms/\suzaku observations and also during the recent \nustar observation. The source is found to exhibit intermittent episodes of X-ray weakness (X-rays diminish more than the UV) as evident by the $\alpha_{ox}$ measurements and captured by \swift. It was not possible to examine the spectrum of this X-ray weak state in detail due to limited data.

On average, the differences between the multi-epoch data can be attributed to changes in the intrinsic power law component, similar to how the rapid variability in 2011 was described.  This is interesting for two reasons. Firstly, on both time scales, the primary variable component is the same. Secondly, there is no strong indication that other spectral components (e.g. disc or absorber properties) respond to changes in the primary component. This may indicate the primary component is located at a rather large distance from the black hole and accretion disc such that the power law component illuminates a large fraction of the disc and changes are smeared out over time (like \Ngc5548, \citealt{Brenneman+2012}).

\subsection{An optically thin and geometrically thick inner disc?}
 \citet{Dobrotka+2017} noted the measured frequency breaks derived from the \kepler and \swift analyses were inconsistent with standard accretion disc time scales and proposed the central disc in \zw229 could be optically thin and geometrically thick. Later studies by \citet{Dobrotka+2019} addressed the Kepler instrument effects and revised the Swift frequency break in \zw229, but the authors conclude that the interpretation proposed in \cite{Dobrotka+2017} remains unchanged. The X-ray spectral modeling in this work is consistent with the \citet{Dobrotka+2017} scenario in that we do not see strong reflection features from the disc. 
 
At odds with this scenario, is that the UV-to-X-ray SED of \zw229 does not appear unusual. The measured $\alpha_{ox}$ and $L/L_{Edd}$ indicate that \zw229 is behaving like a normal, radiative efficient AGN. The continuum and variability are dominated by Comptonisation, which can be described with an optically thick and geometrically thin accretion disc models.

\section{Conclusions}
We have presented a broad band multi-epoch study of the Seyfert 1 galaxy \zw229, combining single-epoch deep X-ray observations and long-term multi-epoch, multi-instrument observations over 9 years. The spectral shape of \zw229 is rather smooth and does not reveal distinct spectral features. The object shows spectral characteristics most comparable to AGN such as \Akn120 and \mrk530. Of interest, is that both rapid (e.g. hours-to-days) and long-term (yearly) variability can be attributed to changes in the brightness and shape of the primary power law component. There is no indication that other X-ray spectral components respond to changes in the power law variability.

Continued monitoring in the X-rays as well as in other wavelengths in combination with the deeper observations from \xmm and \nustar can shed light upon the precise nature of the X-ray variability of this source in an elaborate way. For instance, high-resolution spectroscopy could confirm the absence of absorption, and broadband spectroscopy using simultaneous \nustar and \xmm data will be useful to observe the soft excess and Compton hump which may help determine the correct model.

\section*{Acknowledgements} This research has made use of data and/or software provided by the High Energy Astrophysics Science Archive Research Center (HEASARC), which is a service of the Astrophysics Science Division at NASA/GSFC and the High Energy Astrophysics Division of the Smithsonian Astrophysical Observatory. This work is based on observations obtained with \xmms, an ESA science mission with instruments and contributions directly funded by ESA Member States and NASA. This research has made use of data obtained from the \suzaku satellite, a collaborative mission between the space agencies of Japan (JAXA) and the USA (NASA). This research has made use of the \nustar Data Analysis Software (NuSTARDAS) jointly developed by the ASI ScienceDataCenter (ASDC, Italy) and the California Institute of Technology (USA). This research has made use of the NASA/IPAC Extragalactic Database (NED) which is operated by the Jet Propulsion Laboratory, California Institute of Technology, under contract with the National Aeronautics and Space Administration. This work has made use of data supplied by the UK Swift Science Data Centre at the University of Leicester. LCG acknowledges the support of the Natural Sciences and Engineering Research Council of Canada (NSERC). We thank the anonymous referee for providing useful suggestions to improve the manuscript. The authors also thank Prof. Richard Mushotzky for his comments and suggestions.

\bibliographystyle{mnras}
\bibliography{myref}

\begin{thebibliography}{89}
\expandafter\ifx\csname natexlab\endcsname\relax\def\natexlab#1{#1}\fi

\bibitem[{Adegoke} et~al.(2017){Adegoke}, {Rakshit} \&
  {Mukhopadhyay}]{Adegoke+2017}
{Adegoke} O., {Rakshit} S., {Mukhopadhyay} B., 2017, \mnras, 466, 3951

\bibitem[{Antonucci}(1993)]{Antonucci1993}
{Antonucci} R., 1993, \araa, 31, 473

\bibitem[{Arnaud}(1996)]{Arnaud+1996}
{Arnaud} K.~A., 1996, in { Astronomical Data Analysis Software and Systems
  V\/}, edited by G.~H. {Jacoby}, J.~{Barnes}, vol. 101 of { Astronomical
  Society of the Pacific Conference Series\/}, ~17

\bibitem[{Arnaud} et~al.(1985){Arnaud}, {Branduardi-Raymont}, {Culhane}
  et~al.]{Arnaud+1985}
{Arnaud} K.~A., {Branduardi-Raymont} G., {Culhane} J.~L., et~al., 1985, \mnras,
  217, 105

\bibitem[{Barth} et~al.(2011){Barth}, {Nguyen}, {Malkan} et~al.]{Barth+2011}
{Barth} A.~J., {Nguyen} M.~L., {Malkan} M.~A., et~al., 2011, \apj, 732, 121

\bibitem[{Barthelmy} et~al.(2005){Barthelmy}, {Barbier}, {Cummings}
  et~al.]{Barthelmy+2005}
{Barthelmy} S.~D., {Barbier} L.~M., {Cummings} J.~R., et~al., 2005, \ssr, 120,
  143

\bibitem[{Baumgartner} et~al.(2013){Baumgartner}, {Tueller}, {Markwardt}
  et~al.]{Baumgartner+2013}
{Baumgartner} W.~H., {Tueller} J., {Markwardt} C.~B., et~al., 2013, The
  Astrophysical Journal Supplement Series, 207, 19

\bibitem[{Bonson} \& {Gallo}(2016)]{Bonson+2016}
{Bonson} K., {Gallo} L.~C., 2016, \mnras, 458, 1927

\bibitem[{Bonson} et~al.(2018){Bonson}, {Gallo}, {Wilkins} \&
  {Fabian}]{Bonson+2018}
{Bonson} K., {Gallo} L.~C., {Wilkins} D.~R., {Fabian} A.~C., 2018, \mnras, 477,
  3247

\bibitem[{Brenneman} et~al.(2012){Brenneman}, {Elvis}, {Krongold}, {Liu} \&
  {Mathur}]{Brenneman+2012}
{Brenneman} L.~W., {Elvis} M., {Krongold} Y., {Liu} Y., {Mathur} S., 2012,
  \apj, 744, 13

\bibitem[{Buisson} et~al.(2017){Buisson}, {Lohfink}, {Alston} \&
  {Fabian}]{Buisson+2017}
{Buisson} D.~J.~K., {Lohfink} A.~M., {Alston} W.~N., {Fabian} A.~C., 2017,
  \mnras, 464, 3194

\bibitem[{Burrows} et~al.(2005){Burrows}, {Hill}, {Nousek}
  et~al.]{Burrows+2005}
{Burrows} D.~N., {Hill} J.~E., {Nousek} J.~A., et~al., 2005, \ssr, 120, 165

\bibitem[{Carini} \& {Ryle}(2012)]{Carini+2012}
{Carini} M.~T., {Ryle} W.~T., 2012, \apj, 749, 70

\bibitem[{Cash}(1979)]{Cash1979}
{Cash} W., 1979, \apj, 228, 939

\bibitem[{Chiang} et~al.(2015){Chiang}, {Walton}, {Fabian}, {Wilkins} \&
  {Gallo}]{Chiang+2015}
{Chiang} C.-Y., {Walton} D.~J., {Fabian} A.~C., {Wilkins} D.~R., {Gallo} L.~C.,
  2015, \mnras, 446, 759

\bibitem[{Collinson} et~al.(2017){Collinson}, {Ward}, {Landt}, {Done}, {Elvis}
  \& {McDowell}]{Collinson+2017}
{Collinson} J.~S., {Ward} M.~J., {Landt} H., {Done} C., {Elvis} M., {McDowell}
  J.~C., 2017, \mnras, 465, 358

\bibitem[{Dauser} et~al.(2013){Dauser}, {Garcia}, {Wilms} et~al.]{Dauser+2013}
{Dauser} T., {Garcia} J., {Wilms} J., et~al., 2013, \mnras, 430, 1694

\bibitem[{Dauser} et~al.(2010){Dauser}, {Wilms}, {Reynolds} \&
  {Brenneman}]{Dauser+2010}
{Dauser} T., {Wilms} J., {Reynolds} C.~S., {Brenneman} L.~W., 2010, \mnras,
  409, 1534

\bibitem[{de Vries} et~al.(2003){de Vries}, {Becker} \& {White}]{devries+2003}
{de Vries} W.~H., {Becker} R.~H., {White} R.~L., 2003, \aj, 126, 1217

\bibitem[{den Herder} et~al.(2001){den Herder}, {Brinkman}, {Kahn}
  et~al.]{denHerder+2001}
{den Herder} J.~W., {Brinkman} A.~C., {Kahn} S.~M., et~al., 2001, \aap, 365, L7

\bibitem[{Dobrotka} et~al.(2017){Dobrotka}, {Antonuccio-Delogu} \&
  {Baj{\v{c}}i{\v{c}}{\'a}kov{\'a}}]{Dobrotka+2017}
{Dobrotka} A., {Antonuccio-Delogu} V., {Baj{\v{c}}i{\v{c}}{\'a}kov{\'a}} I.,
  2017, \mnras, 470, 2439

\bibitem[{Dobrotka} et~al.(2019){Dobrotka}, {Bez{\'a}k}, {Revalski} \&
  {Str{\'e}my}]{Dobrotka+2019}
{Dobrotka} A., {Bez{\'a}k} P., {Revalski} M., {Str{\'e}my} M., 2019, \mnras,
  483, 1, 38

\bibitem[{Done} et~al.(2012){Done}, {Davis}, {Jin}, {Blaes} \&
  {Ward}]{Done+2012}
{Done} C., {Davis} S.~W., {Jin} C., {Blaes} O., {Ward} M., 2012, \mnras, 420,
  1848

\bibitem[{Edelson} et~al.(2002){Edelson}, {Turner}, {Pounds}
  et~al.]{Edelson+2002}
{Edelson} R., {Turner} T.~J., {Pounds} K., et~al., 2002, \apj, 568, 610

\bibitem[{Edelson} et~al.(2014){Edelson}, {Vaughan}, {Malkan}
  et~al.]{Edelson+2014}
{Edelson} R., {Vaughan} S., {Malkan} M., et~al., 2014, \apj, 795, 2

\bibitem[{Ehler} et~al.(2018){Ehler}, {Gonzalez} \& {Gallo}]{Ehler+2018}
{Ehler} H.~J.~S., {Gonzalez} A.~G., {Gallo} L.~C., 2018, \mnras, 478, 4214

\bibitem[{Evans} et~al.(2007){Evans}, {Beardmore}, {Page} et~al.]{Evans+2007}
{Evans} P.~A., {Beardmore} A.~P., {Page} K.~L., et~al., 2007, \aap, 469, 379

\bibitem[{Evans} et~al.(2009){Evans}, {Beardmore}, {Page} et~al.]{Evans+2009}
{Evans} P.~A., {Beardmore} A.~P., {Page} K.~L., et~al., 2009, \mnras, 397, 1177

\bibitem[{Fitzpatrick}(1999)]{Fitzpatrick1999}
{Fitzpatrick} E.~L., 1999, \pasp, 111, 63

\bibitem[{Gallant} et~al.(2018){Gallant}, {Gallo} \& {Parker}]{Gallant+2018}
{Gallant} D., {Gallo} L.~C., {Parker} M.~L., 2018, \mnras, 480, 1999

\bibitem[{Gallo}(2006)]{Gallo2006}
{Gallo} L.~C., 2006, \mnras, 368, 479

\bibitem[{Gallo} et~al.(2007{\natexlab{a}}){Gallo}, {Brandt}, {Costantini} \&
  {Fabian}]{Gallo+2007}
{Gallo} L.~C., {Brandt} W.~N., {Costantini} E., {Fabian} A.~C.,
  2007{\natexlab{a}}, \mnras, 377, 1375

\bibitem[{Gallo} et~al.(2007{\natexlab{b}}){Gallo}, {Brandt}, {Costantini} \&
  {Fabian}]{Gallo++2007}
{Gallo} L.~C., {Brandt} W.~N., {Costantini} E., {Fabian} A.~C.,
  2007{\natexlab{b}}, \mnras, 377, 1375

\bibitem[{Gallo} et~al.(2011{\natexlab{a}}){Gallo}, {Grupe}, {Schartel}
  et~al.]{Gallo++2011}
{Gallo} L.~C., {Grupe} D., {Schartel} N., et~al., 2011{\natexlab{a}}, \mnras,
  412, 161

\bibitem[{Gallo} et~al.(2011{\natexlab{b}}){Gallo}, {Miniutti}, {Miller}
  et~al.]{Gallo+2011}
{Gallo} L.~C., {Miniutti} G., {Miller} J.~M., et~al., 2011{\natexlab{b}},
  \mnras, 411, 607

\bibitem[{Gallo} et~al.(2004){Gallo}, {Tanaka}, {Boller}, {Fabian}, {Vaughan}
  \& {Brandt}]{Gallo+2004}
{Gallo} L.~C., {Tanaka} Y., {Boller} T., {Fabian} A.~C., {Vaughan} S., {Brandt}
  W.~N., 2004, \mnras, 353, 1064

\bibitem[{Gallo} et~al.(2015){Gallo}, {Wilkins}, {Bonson} et~al.]{Gallo+2015}
{Gallo} L.~C., {Wilkins} D.~R., {Bonson} K., et~al., 2015, \mnras, 446, 633

\bibitem[{Garc{\'{\i}}a} et~al.(2013){Garc{\'{\i}}a}, {Dauser}, {Reynolds}
  et~al.]{Garcia+2013}
{Garc{\'{\i}}a} J., {Dauser} T., {Reynolds} C.~S., et~al., 2013, \apj, 768, 146

\bibitem[{Garc{\'{\i}}a} et~al.(2011){Garc{\'{\i}}a}, {Kallman} \&
  {Mushotzky}]{Garcia+2011}
{Garc{\'{\i}}a} J., {Kallman} T.~R., {Mushotzky} R.~F., 2011, \apj, 731, 131

\bibitem[{Garc{\'{\i}}a} et~al.(2019){Garc{\'{\i}}a}, {Kara}, {Walton}
  et~al.]{Garcia+2019}
{Garc{\'{\i}}a} J.~A., {Kara} E., {Walton} D., et~al., 2019, \apj, 871, 88

\bibitem[{Gehrels} et~al.(2004){Gehrels}, {Chincarini}, {Giommi}
  et~al.]{Gehrels+2004}
{Gehrels} N., {Chincarini} G., {Giommi} P., et~al., 2004, \apj, 611, 1005

\bibitem[{Goodman} \& {Weare}(2010)]{Goodman+2010}
{Goodman} J., {Weare} J., 2010, Communications in Applied Mathematics and
  Computational Science, Vol.~5, No.~1, p.~65-80, 2010, 5, 65

\bibitem[{Grupe} et~al.(2008){Grupe}, {Komossa}, {Gallo} et~al.]{Grupe+2008}
{Grupe} D., {Komossa} S., {Gallo} L.~C., et~al., 2008, \apj, 681, 982

\bibitem[{Grupe} et~al.(2012){Grupe}, {Komossa}, {Gallo} et~al.]{Grupe+2012}
{Grupe} D., {Komossa} S., {Gallo} L.~C., et~al., 2012, \apjs, 199, 28

\bibitem[{Haardt} \& {Maraschi}(1991)]{Haardt+1991}
{Haardt} F., {Maraschi} L., 1991, \apjl, 380, L51

\bibitem[{Haardt} \& {Maraschi}(1993)]{Haardt+1993}
{Haardt} F., {Maraschi} L., 1993, \apj, 413, 507

\bibitem[{Harrison} et~al.(2013){Harrison}, {Craig}, {Christensen}
  et~al.]{Harrison+2013}
{Harrison} F.~A., {Craig} W.~W., {Christensen} F.~E., et~al., 2013, \apj, 770,
  103

\bibitem[{Jansen} et~al.(2001){Jansen}, {Lumb}, {Altieri} et~al.]{Jansen+2001}
{Jansen} F., {Lumb} D., {Altieri} B., et~al., 2001, \aap, 365, L1

\bibitem[{Kaastra} \& {Bleeker}(2016)]{Kaastra+2016}
{Kaastra} J.~S., {Bleeker} J.~A.~M., 2016, \aap, 587, A151

\bibitem[{Kalberla} et~al.(2005){Kalberla}, {Burton}, {Hartmann}
  et~al.]{Kalberla+2005}
{Kalberla} P.~M.~W., {Burton} W.~B., {Hartmann} D., et~al., 2005, \aap, 440,
  775

\bibitem[{Kasliwal} et~al.(2015){Kasliwal}, {Vogeley} \&
  {Richards}]{Kasliwal+2015}
{Kasliwal} V.~P., {Vogeley} M.~S., {Richards} G.~T., 2015, \mnras, 451, 4328

\bibitem[{Kasliwal} et~al.(2017){Kasliwal}, {Vogeley} \&
  {Richards}]{Kasliwal+2017}
{Kasliwal} V.~P., {Vogeley} M.~S., {Richards} G.~T., 2017, \mnras, 470, 3027

\bibitem[{Kelly} et~al.(2014){Kelly}, {Becker}, {Sobolewska}, {Siemiginowska}
  \& {Uttley}]{Kelly+2014}
{Kelly} B.~C., {Becker} A.~C., {Sobolewska} M., {Siemiginowska} A., {Uttley}
  P., 2014, \apj, 788, 33

\bibitem[{Lohfink} et~al.(2016){Lohfink}, {Reynolds}, {Pinto}
  et~al.]{Lohfink+2016}
{Lohfink} A.~M., {Reynolds} C.~S., {Pinto} C., et~al., 2016, \apj, 821, 1, 11

\bibitem[{Malzac} et~al.(2006){Malzac}, {Petrucci}, {Jourdain}
  et~al.]{Malzac+2006}
{Malzac} J., {Petrucci} P.~O., {Jourdain} E., et~al., 2006, \aap, 448, 1125

\bibitem[{Mason} et~al.(2001){Mason}, {Breeveld}, {Much} et~al.]{Mason+2001}
{Mason} K.~O., {Breeveld} A., {Much} R., et~al., 2001, \aap, 365, L36

\bibitem[{Matt} et~al.(2014){Matt}, {Marinucci}, {Guainazzi} et~al.]{Matt+2014}
{Matt} G., {Marinucci} A., {Guainazzi} M., et~al., 2014, \mnras, 439, 3016

\bibitem[{Middei} et~al.(2019){Middei}, {Bianchi}, {Petrucci}
  et~al.]{Middei+2019}
{Middei} R., {Bianchi} S., {Petrucci} P.-O., et~al., 2019, \mnras, 483, 4695

\bibitem[{Miller}(2007)]{Miller2007}
{Miller} J.~M., 2007, \araa, 45, 441

\bibitem[{Miller} et~al.(2009){Miller}, {Turner} \& {Reeves}]{Miller+2009}
{Miller} L., {Turner} T.~J., {Reeves} J.~N., 2009, \mnras, 399, L69

\bibitem[{Miniutti} et~al.(2009){Miniutti}, {Fabian}, {Brandt}, {Gallo} \&
  {Boller}]{Miniutti+2009}
{Miniutti} G., {Fabian} A.~C., {Brandt} W.~N., {Gallo} L.~C., {Boller} T.,
  2009, \mnras, 396, L85

\bibitem[{Mitsuda} et~al.(2007){Mitsuda}, {Bautz}, {Inoue}
  et~al.]{Mitsuda+2007}
{Mitsuda} K., {Bautz} M., {Inoue} H., et~al., 2007, \pasj, 59, S1

\bibitem[{Mushotzky} et~al.(2011){Mushotzky}, {Edelson}, {Baumgartner} \&
  {Gandhi}]{Mushotzky+2011}
{Mushotzky} R.~F., {Edelson} R., {Baumgartner} W., {Gandhi} P., 2011, \apj,
  743, L12

\bibitem[{Nowak} et~al.(2011){Nowak}, {Hanke}, {Trowbridge} et~al.]{Nowak+2011}
{Nowak} M.~A., {Hanke} M., {Trowbridge} S.~N., et~al., 2011, \apj, 728, 13

\bibitem[{Parker} et~al.(2015){Parker}, {Fabian}, {Matt} et~al.]{Parker+2015}
{Parker} M.~L., {Fabian} A.~C., {Matt} G., et~al., 2015, \mnras, 447, 72

\bibitem[{Parker} et~al.(2014){Parker}, {Marinucci}, {Brenneman}
  et~al.]{Parker+2014a}
{Parker} M.~L., {Marinucci} A., {Brenneman} L., et~al., 2014, \mnras, 437, 721

\bibitem[{Petrucci} et~al.(2018){Petrucci}, {Ursini}, {De Rosa}
  et~al.]{Petrucci+2018}
{Petrucci} P.-O., {Ursini} F., {De Rosa} A., et~al., 2018, \aap, 611, A59

\bibitem[{Ponti} et~al.(2012){Ponti}, {Papadakis}, {Bianchi}
  et~al.]{Ponti+2012}
{Ponti} G., {Papadakis} I., {Bianchi} S., et~al., 2012, \aap, 542, A83

\bibitem[{Porquet} et~al.(2018){Porquet}, {Reeves}, {Matt}
  et~al.]{Porquet+2018}
{Porquet} D., {Reeves} J.~N., {Matt} G., et~al., 2018, \aap, 609, A42

\bibitem[{Pounds} et~al.(2004){Pounds}, {Reeves}, {King} \&
  {Page}]{Pounds+2004}
{Pounds} K.~A., {Reeves} J.~N., {King} A.~R., {Page} K.~L., 2004, \mnras, 350,
  1, 10

\bibitem[{Pravdo} et~al.(1981){Pravdo}, {Nugent}, {Nousek}, {Jensen}, {Wilson}
  \& {Becker}]{Pravdo+1981}
{Pravdo} S.~H., {Nugent} J.~J., {Nousek} J.~A., {Jensen} K., {Wilson} A.~S.,
  {Becker} R.~H., 1981, \apj, 251, 501

\bibitem[{Reeves} et~al.(2008){Reeves}, {Done}, {Pounds} et~al.]{Reeves+2008}
{Reeves} J., {Done} C., {Pounds} K., et~al., 2008, \mnras, 385, L108

\bibitem[{Roming} et~al.(2005){Roming}, {Kennedy}, {Mason} et~al.]{Roming+2005}
{Roming} P.~W.~A., {Kennedy} T.~E., {Mason} K.~O., et~al., 2005, \ssr, 120, 95

\bibitem[{Ross} \& {Fabian}(2005)]{Ross+2005}
{Ross} R.~R., {Fabian} A.~C., 2005, \mnras, 358, 211

\bibitem[{Schartel} et~al.(2010){Schartel}, {Rodr{\'{\i}}guez-Pascual},
  {Santos-Lle{\'o}}, {Jim{\'e}nez-Bail{\'o}n}, {Ballo} \&
  {Piconcelli}]{Schartel+2010}
{Schartel} N., {Rodr{\'{\i}}guez-Pascual} P.~M., {Santos-Lle{\'o}} M.,
  {Jim{\'e}nez-Bail{\'o}n} E., {Ballo} L., {Piconcelli} E., 2010, \aap, 512,
  A75

\bibitem[{Singh} et~al.(1985){Singh}, {Garmire} \& {Nousek}]{Singh+1985}
{Singh} K.~P., {Garmire} G.~P., {Nousek} J., 1985, \apj, 297, 633

\bibitem[{Smith} et~al.(2018){Smith}, {Mushotzky}, {Boyd}, {Malkan}, {Howell}
  \& {Gelino}]{Smith+2018}
{Smith} K.~L., {Mushotzky} R.~F., {Boyd} P.~T., {Malkan} M., {Howell} S.~B.,
  {Gelino} D.~M., 2018, \apj, 857, 141

\bibitem[{Strateva} et~al.(2005){Strateva}, {Brandt}, {Schneider}, {Vanden
  Berk} \& {Vignali}]{Strateva+2005}
{Strateva} I.~V., {Brandt} W.~N., {Schneider} D.~P., {Vanden Berk} D.~G.,
  {Vignali} C., 2005, \aj, 130, 387

\bibitem[{Str{\"u}der} et~al.(2001){Str{\"u}der}, {Briel}, {Dennerl}
  et~al.]{Struder+2001}
{Str{\"u}der} L., {Briel} U., {Dennerl} K., et~al., 2001, \aap, 365, L18

\bibitem[{Tanaka} et~al.(2004){Tanaka}, {Boller}, {Gallo}, {Keil} \&
  {Ueda}]{Tanaka+2004}
{Tanaka} Y., {Boller} T., {Gallo} L., {Keil} R., {Ueda} Y., 2004, \pasj, 56, L9

\bibitem[{Tananbaum} et~al.(1979){Tananbaum}, {Avni}, {Branduardi}
  et~al.]{Tananbaum+1979}
{Tananbaum} H., {Avni} Y., {Branduardi} G., et~al., 1979, \apjl, 234, L9

\bibitem[{Thomas} et~al.(2016){Thomas}, {Groves}, {Sutherland}, {Dopita},
  {Kewley} \& {Jin}]{Thomas+2016}
{Thomas} A.~D., {Groves} B.~A., {Sutherland} R.~S., {Dopita} M.~A., {Kewley}
  L.~J., {Jin} C., 2016, \apj, 833, 266

\bibitem[{Turner} \& {Miller}(2009)]{Turner+2009}
{Turner} T.~J., {Miller} L., 2009, \aapr, 17, 47

\bibitem[{Urry} \& {Padovani}(1995)]{Urry+1995}
{Urry} C.~M., {Padovani} P., 1995, \pasp, 107, 803

\bibitem[{Vagnetti} et~al.(2011){Vagnetti}, {Turriziani} \&
  {Trevese}]{Vagnetti+2011}
{Vagnetti} F., {Turriziani} S., {Trevese} D., 2011, \aap, 536, A84

\bibitem[{Vaughan} et~al.(2004){Vaughan}, {Fabian}, {Ballantyne}, {De Rosa},
  {Piro} \& {Matt}]{Vaughan+2004}
{Vaughan} S., {Fabian} A.~C., {Ballantyne} D.~R., {De Rosa} A., {Piro} L.,
  {Matt} G., 2004, \mnras, 351, 193

\bibitem[Wilkins et~al.(2017)Wilkins, Gallo, Silva, Costantini, Brandt \&
  Kriss]{Wilkins+2017}
Wilkins D., Gallo L., Silva C., Costantini E., Brandt W., Kriss G., 2017,
  Monthly Notices of the Royal Astronomical Society, 471, 4, 4436

\bibitem[{Wilkins} \& {Fabian}(2011)]{Wilkins+2011}
{Wilkins} D.~R., {Fabian} A.~C., 2011, \mnras, 414, 1269

\bibitem[{Wilkins} \& {Gallo}(2015)]{Wilkins+2015}
{Wilkins} D.~R., {Gallo} L.~C., 2015, \mnras, 448, 703

\end{thebibliography}

\end{document}